\documentclass[sigconf,balance=false,nonacm]{acmart}
\usepackage[utf8]{inputenc}
\usepackage[normalem]{ulem}
\usepackage{amsmath}
\usepackage[ruled, linesnumbered]{algorithm2e}
\usepackage{xcolor}
\usepackage{paralist}
\usepackage{graphicx}
\usepackage{xurl}
\usepackage{pifont}
\usepackage{threeparttable}
\usepackage{comment}
\usepackage{lscape}
\usepackage{caption}
\usepackage{subcaption}
\usepackage{float}
\usepackage{url}
\usepackage{multirow}

\mathchardef\mhyphen="2D
\newcommand{\cmark}{\ding{51}}%
\newcommand{\xmark}{\ding{55}}%
\SetKwInput{KwInput}{Input}
\SetKwInput{KwOutput}{Output}  
\SetKw{KwVariable}{Variable}  
\SetKwProg{Fn}{Function}{:}{}
\SetKw{Continue}{continue}
\SetKw{Goto}{goto}

\begin{document}
\title{Watching your call: Breaking VoLTE Privacy in LTE/5G Networks}

\author{Zishuai Cheng}
\affiliation{%
  \institution{Beijing University of Posts and Telecommunications}
  \streetaddress{10 Xitucheng Rd}
  \city{Haidian Qu}
  \state{Beijing}
  \country{China}}
\email{chengzishuai@bupt.edu.cn}

\author{Mihai Ordean}
\affiliation{%
  \institution{University of Birmingham}
  \streetaddress{}
  \city{Birmingham}
  \country{UK}}
\email{m.ordean@bham.ac.uk}

\author{Flavio D. Garcia}
\affiliation{%
  \institution{University of Birmingham}
  \streetaddress{}
  \city{Birmingham}
  \country{UK}}
\email{f.garcia@bham.ac.uk}

\author{Baojiang Cui}
\affiliation{%
  \institution{Beijing University of Posts and Telecommunications}
  \streetaddress{10 Xitucheng Rd}
  \city{Haidian Qu}
  \state{Beijing}
  \country{China}}
\email{cuibj@bupt.edu.cn}

\author{Dominik Rys}
\affiliation{%
  \institution{University of Birmingham}
  \streetaddress{}
  \city{Birmingham}
  \country{UK}}
\email{dominik.j.rys@gmail.com}

\begin{abstract}
Voice over LTE (VoLTE) and Voice over NR (VoNR), are two similar technologies that have been widely deployed by operators to provide a better calling experience in LTE and 5G networks, respectively. The VoLTE/NR protocols rely on the security features of the underlying LTE/5G network to protect users' privacy such that nobody can monitor calls and learn details about call times, duration, and direction. In this paper, we introduce a new privacy attack which enables adversaries to analyse encrypted LTE/5G traffic and recover any VoLTE/NR call details. We achieve this by implementing a novel mobile-relay adversary which is able to remain undetected by using an improved physical layer parameter guessing procedure. This adversary facilitates the recovery of encrypted configuration messages exchanged between victim devices and the mobile network. We further propose an identity mapping method which enables our mobile-relay adversary to link a victim's network identifiers to the phone number efficiently, requiring a single VoLTE protocol message. We evaluate the real-world performance of our attacks using four modern commercial off-the-shelf phones and two representative, commercial network carriers. We collect over 60 hours of traffic between the phones and the mobile networks and execute 160 VoLTE calls, which we use to successfully identify patterns in the physical layer parameter allocation and in VoLTE traffic, respectively. Our real-world experiments show that our mobile-relay works as expected in all test cases, and the VoLTE activity logs recovered describe the actual communication with 100\% accuracy. Finally, we show that we can link network identifiers such as International Mobile Subscriber Identities (IMSI), Subscriber Concealed Identifiers (SUCI) and/or Globally Unique Temporary Identifiers (GUTI) to phone numbers while remaining undetected by the victim.
\end{abstract}


\maketitle

\section{Introduction}\label{introduction}
Mobile communication technologies are used by billions of people around the world in their daily lives. While the latest mobile communication technology is 5G, the previous generation technology 4G, sometimes named Long-Term Evolution (LTE), still dominates the market~\cite{the-mobile-economy}. The core elements in both LTE and 5G are: the User Equipment (UE), the cell tower known as E-UTRAN Node B (eNodeB) in LTE or Next Generation Node B (gNodeB) in 5G, and the core network known as Evolved Packet Core (EPC) in LTE. The UE is a user device, such as a mobile phone, which contains a Universal Subscriber Identity Module (USIM) able to perform cryptographic operations for authentication purposes using a cryptographic key pre-shared with the carrier network. The USIM module either stores, or is able to generate, unique values that UEs used to identify themselves to the network. These identifiers fall into two categories: permanent identifiers such as IMSI and temporary identifiers such as SUCI. Given that UE's communication with the eNodeB is done over the radio, the temporary identifiers along with integrity protection and encryption mechanisms are used to provide confidentiality and protect users' privacy by preventing unauthorised access to data logs, call logs or conversation activities. 

The Voice over IP (VoIP) technology has been added to mobile communication with LTE in order to support voice communication in packet-switched exclusive networks\footnote{In 2G and 3G networks voice is transferred using dedicated analogue channels} and to provide a better call experience (e.g., lower setup time and lower latency). Known as VoLTE in LTE or Voice over NR in 5G, it uses an IP Multimedia Subsystem (IMS) which is deployed out of the core network, but which is still controlled by the network carrier in order to facilitate payment for the service. As VoLTE/NR services in LTE/5G transfer signalling and voice data over-the-air, an adversary could observe the connections and the traffic exchanges if protections are not deployed appropriately. Given the similarities between VoLTE and VoNR, throughout the paper we will refer to both as VoLTE and make the distinction where required.

Unfortunately, recent studies reveal that the data exchanged between the UEs and the eNodeB, i.e. the cell tower, is not well protected. Radio signal \textit{overpowering} for the purposes of data overwriting on the physical layer (e.g., Layer 1) has been shown to be effective at influencing the data received by UEs~\cite{yang2019hiding}. This can further allow adversaries to launch Denial of Service (DoS) attacks and collect victim identifiers, such as IMSIs~\cite{arxiv.2106.05039}.

Furthermore, Layer 2 attacks have also been proven effective by Rupprecht et al. which proposes a relay type adversary which forwards data between victim UEs and a commercial eNodeB~\cite{rupprecht-breaking-2019}. This relay attacker is significantly different from the cellular repeater which is commonly used to boost the cellular signals, as the relay first picks up and demodulates the radio signal to bits and then modulates bits and transmits to reception using proper radio resources (e.g., carrier frequency and transmission time), whereas the repeater is only amplifying the power of the signals and functions only on the physical layer. Several other attacks have been proposed which are able to tamper, recover or \textit{fingerprint} the data transmitted over-the-air. Tampering Internet data, recovering voice data and \textit{impersonating} attacks are proposed by Rupprecht et al.~\cite{rupprecht-breaking-2019, rupprecht-call-2020, rupprecht2020imp4gt}. In contrast, several weaker attackers~\cite{kohls-lost-2019, 277230} are proposed to \textit{fingerprint} victim's data, which can monitor victims' activities about browsing websites and watching videos. These attacks significantly break the privacy requirements of LTE/5G which requires that no one is able to monitor users' activities.

In this paper, we present the first study focused on the analysis of encrypted VoLTE traffic consisting of both signalling data, the VoLTE messages exchanged between a UE and the IMS, and voice data, representing voice activities observed in windows of 20ms. These insights allow us to develop means for monitoring specific VoLTE activities enabling us to learn conversation states of targeted victims and their relationship with other victims, while being located in one or more areas, e.g., victim A calls victim B at a time T and talks for the majority of the conversation.

\subsection{Contributions}
We develop, deploy and test a novel LTE/5G mobile-relay, based on open source software and commercial off-the-shelf (COTS) hardware, significantly improving on existing work~\cite{rupprecht-breaking-2019}. Using this relay, which allows us to intercept and monitor connections between victim UEs and commercial eNodeBs, in this paper, we show:

\begin{enumerate}
    \item The first privacy attack that targets encrypted LTE and 5G-SA traffic to extract VoLTE activity logs which describe call times, duration, and speaker direction for users in mobile networks.
    \item A novel and efficient identity mapping method which links phone numbers to LTE and 5G-SA network identifiers. Our attack is completely undetectable when used to link phone numbers to temporary identifiers, and has minimal protocol interference when linking them to permanent ones.
    \item Several physical layer improvements to the mobile-relay adversary, which greatly improve the effectiveness of this attacker.
\end{enumerate}

We evaluate the feasibility of our contributions above by testing them using four COTS phones and two major commercial carriers. 

\section{Preliminaries}\label{preliminaries}
In this section, we give an overview of the main, relevant technologies investigated in this paper.

\subsection{LTE/5G network communication}
\begin{figure}[t]                      
    \centering
    \includegraphics[width=\columnwidth,trim={10 10 10 10},clip]{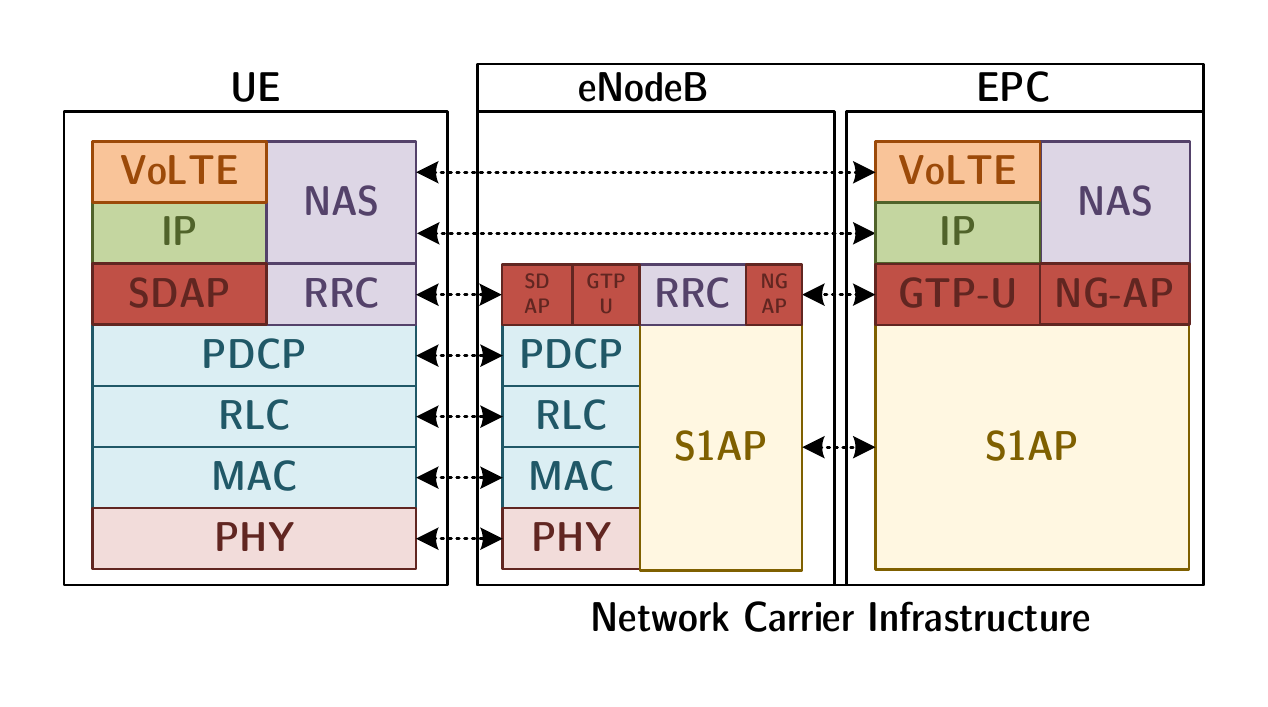}
    \caption{Overview of 5G/LTE radio access network architecture. Components marked in red are 5G specific and do not contain any security-related features. Some 5G sub-layers have been omitted for brevity.}
    \label{fig:lte_overv}
\end{figure}

From a high-level view, as previously stated, LTE and 5G networks consist of three main components: the user equipment, the eNodeB, and the evolved packet core. The EPC contains all the software and hardware components that provide necessary functionalities such as data and voice communication services between UEs, authentication and billing. Communication between these three entities is done differently, based on the requirements and location, as shown in Fig.~\ref{fig:lte_overv}. Given that both the eNodeB and the EPC are components of the carrier network's infrastructure, the security here is mostly ensured through physical means such as having wired connections to transport the S1 Application Protocol (S1AP) protocol messages. The radio link between the UE and the eNodeB, on the other hand, is susceptible to interception and interference from any number of actors and, therefore, has more security and reliability features built-in. While an attacker that wants to target specific services running inside the EPC can consider both these links as viable, the radio link provides a significantly more accessible and less tamper-evident entry point, if the security features can be circumvented. We continue by presenting a brief overview of the protocol layers used on the radio access link, which is the one targeted by our mobile-relay adversary.

\vspace{5pt}
\noindent\textbf{LTE/5G radio access architecture.} LTE and 5G protocols use a wide range of frequency bands located from 1GHz to 6GHz and mmWaves (30–300GHz) in the new 5G standard. Data modulation and encoding on these frequencies are handled at the physical layer (PHY) of the protocol and can be done using Frequency-Division Duplex (FDD), Time-Division Duplexing (TDD) or FDD Supplemental Downlink (SDL). The Medium Access Control (MAC) layer is the first logical layer of the protocol stack and is responsible for exchanging measurements and parameters such as channel quality indicators and modulation schemes, which are used to adjust the PHY layer and ensure the best quality of communication. The Radio Link Control (RLC) layer sits above the MAC layer and provides necessary error correction, segmentation and broadcast capabilities to the layers above. The Packet Data Convergence Protocol (PDCP) is the layer which handles cryptographic keys and provides encryption and integrity protection to the layers above. This is particularly important in an adversarial setting because all traffic encapsulated in PDCP packets (such as VoLTE traffic) is at least encrypted. Finally, the network layer is formed of three sub-layers: (1) the Radio Resource Control (RRC) sub-layer which connects the UE to the eNodeB and facilitates the exchange of configuration messages for the lower layers, including MAC and PHY layers, using encrypted PDCP messages; (2) the Non-Access Stratum (NAS) sub-layer which connects the UE to the EPC through RRC messages initially and then S1AP messages, and is responsible for authentication and mobility within the network, and (3) the IP (or user-plane (UP)) sub-layer which connects the UE to the core network through encrypted PDCP packets and is responsible for providing user services such as Internet access or VoLTE.

\subsection{Mobile-relay adversarial node}
\begin{figure}                          
    \centering
    \includegraphics[width=\columnwidth]{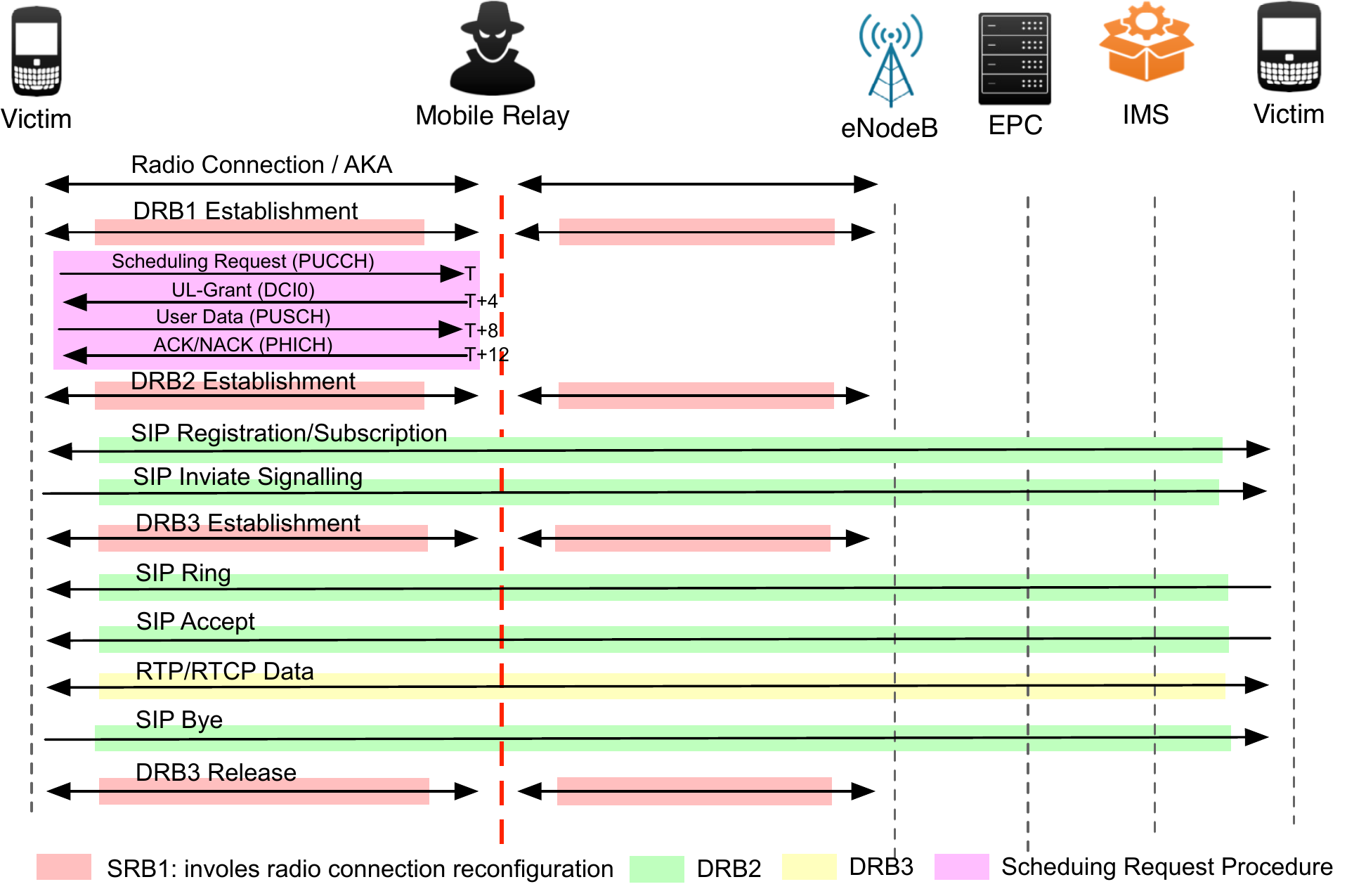}
    \caption{VoLTE protocol message diagram. The mobile-relay adversary is located between the victim UE(s) and commercial eNodeB. The relay maintains two independent physical layer radio connections and forwards encrypted PDCP layer traffic between the UE(s) and the eNodeB. \textit{Scheduling Request} procedure outlines the method in which UE requests an uplink transmission resource to transmit data, from the mobile-relay. Every other type of traffic is normally encrypted by the UE or the eNodeB and thus forwarded without alterations.}
    \label{fig:framework}
\end{figure}

We design and build a mobile-relay adversary that is positioned between the victim UE and the eNodeB and behaves as a Man-in-the-Middle attacker. This relay adversary maintains two independent physical layer radio connections: one to connect to victim UE(s), and another with the eNodeB (see Fig.~\ref{fig:framework}) similar to the one proposed in~\cite{rupprecht-breaking-2019}. As, these two physical connections are separately maintained, and thus direct traffic forwarding is only possible at higher layers, e.g., PDCP and RRC (see Fig.~\ref{fig:lte_overv}).

Maintaining the connections, however, is challenging because after the initial connection stages, all subsequent physical layer configuration parameters are exchanged using encrypted RRC messages. This forces the attacker to continuously guess the physical layer parameters in order to maintain its radio connections alive. We discuss our improvements and how we reliably address the problems in Section~\ref{breaking_volte_privacy}.

\subsection{VoLTE service}
In this section, we describe the VoLTE service following IMS deployed in the carrier's network, the radio bearers used to transmit VoLTE traffic, related protocols and the VoLTE client application specifics provisioned on UEs.

\vspace{5pt}
\noindent\textbf{IMS.} IMS is a standalone system for providing IP multimedia services, session management and media control. An important component of IMS is the Proxy Call Session Control Function (P-CSCF) entity, which directly interacts with VoLTE clients. The Session Initiation Protocol (SIP) together with the Real-time Transport Protocol (RTP) and the RTP Control Protocol (RTCP) are used in VoLTE to manage call sessions, deliver audio data and report transmission state, respectively. In this work, we exploit leaks from these protocols in order to reveal details about connections that should be protected, thus breaking the privacy of VoLTE.

\vspace{5pt}
\noindent\textbf{Radio bearers.} 3GPP assigns different services with different transmission priorities indicated by QoS Class Identifier (QCI) to improve user experience~\cite{3gpp-23203}. To this end, LTE sets up an Evolved Packet-switched System (EPS) Bearer between UE and Packet Data Network Gateway (P-GW) for each QCI, and identifies these bearers with Data Radio Bearer (DRB) ids. Each DRB is associated with a Logical Channel ID (LCID) at the MAC layer. When using VoLTE, SIP packets are transmitted on DRB2 using LCID 4 and QCI 5, while RTP packets use DRB3, LCID 5 and QCI 1. RTCP packets can be transmitted either on DRB2 or on DRB3 which depends on the carriers' configuration. To further reduce the VoLTE bandwidth, 3GPP introduces Robust Header Compression (ROHC) to squeeze bulky protocol headers (e.g., IPv6 header, UDP header, RTP header) to exactly 3 bytes~\cite{rfc-3095, 3gpp-36323}. In this work, we mostly focus on the traffic transmitted on DRB2 and DRB3 which is related to VoLTE activities.

\vspace{5pt}
\noindent\textbf{SIP/RTP/RTCP.} As shown in Fig.~\ref{fig:framework}, after DRB2 is established, the UE registers to the IMS and then subscribes to events from the IMS (e.g., incoming call events). When a call is accepted, as a consequence of receiving an \textit{Invite} message from a caller, a DRB3 bearer is established to prepare for the transmission of audio data. The audio data is sent using RTP packets. The call session is terminated when a \textit{Bye} message is sent. This results in the immediate release of DRB3. During the conversation, two types of RTP packets can be sent, one contains the encoded audio frame, and the other contains a single \textit{Comfort Noise} frame. The first type of packet is transferred every 20ms while the latter is transferred every 160ms. And the size of \textit{Comfort Noise} frame is 6 bytes which is much smaller than other frames~\cite{3gpp-26071, 3gpp-26201, 3gpp-26102}. This frame, however, is only sent when the Voice Activity Detector (VAD) identifies that the speaker has not spoken in the last sampling period, the purpose being to save the bandwidth and battery life. The use of \textit{Comfort Noise} frame allows us to monitor the victim's voice activity with a high granularity by analysing uplink and downlink bit-rate separately. We detail this more in Section~\ref{call-signalling}.

\noindent\textbf{VoLTE client.} VoLTE client is usually part of the software stack running on COTS phones, however, and uses the aforementioned public protocols (e.g., SIP, RTP) to provide VoLTE services. This client connects to the carrier's IMS and encodes the user's operations as specific SIP messages based on predefined templates. These templates are only relevant to specific vendor implementations but, based on our observations, they are static. This enables an attacker to compile VoLTE signalling logs (e.g., SIP messages) by evaluating the communication characteristics of the traffic.

\section{Breaking privacy using VoLTE}\label{breaking_volte_privacy}
The process of breaking users' privacy using VoLTE (or VoNR in 5G) mainly involves recovering the VoLTE activity logs belonging to the victim, including both signalling and voice logs. We refer to \textit{signalling logs} as the part of the traffic comprised of SIP messages exchanged between the victim UE and the carrier's IMS. Conversely, by \textit{voice logs} we refer exclusively to the voice packets exchanged between victims. By leveraging these self-computed logs we can reveal the links between the anonymised network identifiers (e.g., SUCI, Temporary IMSI (T-IMSI)) and real victim identities, i.e. phone numbers. To this end, we use a mobile-relay to collect victim identifiers and the encrypted VoLTE traffic exchanged between UEs and the IMS. We exploit the static nature of VoLTE data to extract meaningful information from the encrypted traffic. In the following, we introduce our threat model followed by descriptions of our attacks.

\subsection{Threat Model}
We begin our threat model analysis by introducing the main goals of the adversary as: (1) \textit{data collection}, which represents the adversary's goal to stealthily collect relevant data, such as plaintext network configuration parameters, identifiers and encrypted traffic; (2) \textit{VoLTE data analysis}, the goal of successfully processing the collected traffic for the purposes of extracting meaningful information such as VoLTE logs; and (3) \textit{real-world identity mapping}, the goal of associating collected traffic to real-world victims identified through their phone numbers.

Next, we map these against three types of adversaries sorted from weakest to strongest as follows. First, our weakest adversary is a completely \textit{passive adversary} located between the UE and the network provider. This adversary is able to achieve both the \textit{data collection} and \textit{traffic analysis} goals. This is a similar attacker model to the one proposed by Rupprecht et al.~\cite{rupprecht-breaking-2019}, which is able to redirect Radio Frequency (RF) domain data flows through an attacker controlled node, however, we expand the capabilities of this with additional data processing at the radio communication level greatly improving stealthiness and reliability. This adversary is able to observe both uplink and downlink radio communication data between the UE and the network at the physical layer. While this attack does require the adversary to initiate a standard UE attach procedure, we maintain that this attacker can be seen as passive as it remains silent with respect to the data flow, the attach procedure is indistinguishable from a legitimate one, and the attacker does not have access to any cryptographic material belonging either to the network or the UE. We also highlight that, from a functional point of view, RF data redirection is not a necessary requirement and attacker models, such as the fully passive one proposed by Kotuliak et al.~\cite{kotuliak2022ltrack}, would be equally efficient. 

Our next two attacker models deal with the problem of \textit{real-world identity mapping}, which requires some form of data exchange between the attacker and the victim. As such, our mid-strength model is a \textit{passive adversary with call capabilities}. We require that this attacker has knowledge of the victim's phone number and can initiate VoLTE calls identical to a standard UE. Additional UE functionality however is not required. This attacker can remain undetectable given that it fully obeys protocols by only interacting with the victim using stranded functionally.

Finally, our strongest adversary is an \textit{active adversary} which is able to initiate calls and perform modifications to the data exchanged between the UE and the network. This adversary, however, still does not have any access to cryptographic materials belonging to the network or the UE. Due to its ability to modify traffic, this attacker is potentially detectable. We discuss the challenges of detecting this attack in Section~\ref{exp imsi catcher detection}.

We implement our attacks, using COTS UEs, software-defined radio (SDR) devices, and a modified version of the open-source srsRAN mobile communication software stack~\cite{srsran}.

\subsection{Obtaining physical layer parameters}
\label{volte-related-packets}
\begin{figure}[htb]
    \centering
    \includegraphics[width=0.9\columnwidth]{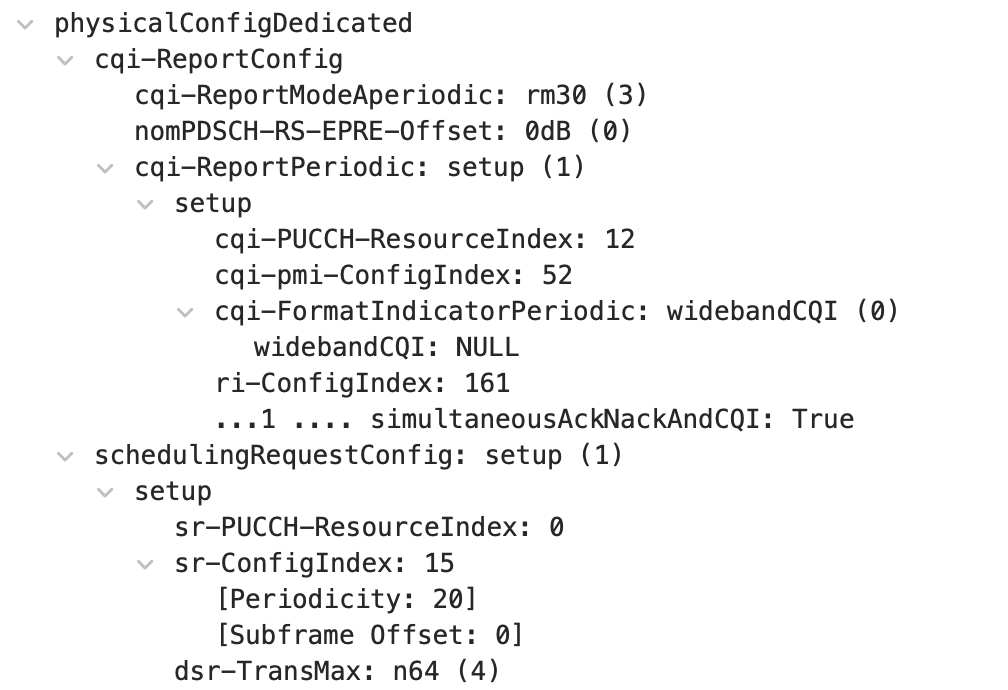}
    \caption{An example of physical layer configuration indicated by eNodeB. \textit{cqi-ReportConfig} and \textit{schedulingRequestConfig} are important to indicate the time (e.g., sub-frame in time domain) and frequency (e.g., sub-carrier in frequency domain) to send \textit{CQI} and \textit{SR} messages. These configuration messages are encrypted and parameter values are unknown to the adversary.}\label{fig:physical_config_dedicated}
\end{figure}

The physical layer of a 5G/LTE network, in the normal mode of operation, allocates radio resources, i.e. the smallest data units used by mobile networks, dynamically in order to avoid interference and exploit the bandwidth efficiently. This process begins when a UE sends a \textit{Scheduling Request (SR)} message to the eNodeB component of the network to request an Uplink Shared Channel (UL-SCH) resource for uplink data transmissions. After the connection is established, the UE needs to periodically report to the eNodeB the channel quality using \textit{Channel Quality Indicator (CQI)} messages, which affect the Modulation and Coding Scheme (MCS) used between the two. In case the UE fails repeatedly to send \textit{SR} or \textit{CQI} reports, the radio connection is terminated~\cite{3gpp-36213,3gpp-36321}. Due to reasons related to signal changes, optimal resource allocation, establish/release EPS bearer, and/or bandwidth efficiency, RLC, MAC, and PHY parameters can be updated by the eNodeB through \textit{RRCConnectionReconfiguration} messages. While RLC and MAC parameters remain fairly static over the course of a connection, physical layer parameters, which are used to orchestrate the all connected subscribers on the radio spectrum, are frequently adjusted. Without knowledge of these, the adversary is unable to maintain the connection between the victim and the eNodeB as it cannot allocate or use the correct radio resources. Furthermore, when such a situation is encountered, the radio connection is immediately released and is followed by a new random access procedure. An example of these parameters is shown in Fig.~\ref{fig:physical_config_dedicated} where the \textit{physicalConfigDedicated} entry specifies the physical layer parameters. The two most important entities are \textit{schedulingRequestConfig} which is responsible for requesting radio resources to be used for sending uplink data (i.e. via the Physical Uplink Shared Channel (PU-SCH)), and \textit{cqi-ReportConfig} which instructs on the type of MCS the eNodeB should use. 

Given the location of our mobile-relay, the attacker can continuously monitor the communication stream and look for encrypted \textit{RRCConnectionReconfiguration} messages\footnote{The adversary cannot locate this message by examining the context because messages are encrypted, but the message can still be identified by examining its length and position in the protocol sequence.}. When such a message is detected, the eNodeB interface of mobile-relay opens up all proper radio resources, i.e. all slots in the time domain and sub-carriers in the frequency domain, and then waits for the victim UE to use one of them. The mobile-relay continuously monitors the radio resources used by the victim UE to transmit uplink data until the mobile-relay obtains the physical layer parameters, then the mobile-relay applies these parameters on both eNodeB and UE interface and removes redundant radio resources. We describe the details of guessing \textit{schedulingRequestConfig} and \textit{cqi-ReportConfig} as follows.

\vspace{5pt}
\noindent\textbf{Recovering \textit{schedulingRequestConfig} parameters.} After receiving an \textit{Scheduling Request (SR)} message from a UE at a time $T$, the eNodeB assigns this UE a radio resource for transmitting uplink data. This assignment is communicated to the UE via \textit{Uplink Grant (UL-Grant)} at time $T+4ms$. If the UE does not receive \textit{UL-Grant} response at $T+4ms$, it will send another \textit{SR} request at the next available period. This process can be repeated until it reaches the maximum re-transmission threshold allowed, which is indicated by the \textit{dsr-TransMax} parameter. The process is shown in Fig.~\ref{fig:framework}.

In order to compute \textit{sr-ConfigIndex} and \textit{sr-PUCCH-ResourceIndex} we proceed as follows. The process begins with the mobile-relay listening for a \textit{RRCConnectionReconfiguration} message sent by the commercial eNodeB. When this is observed, the relay starts monitoring all slots in the time domain and all sub-carriers in the frequency domain. Then, using the first \textit{SR} message intercepted, the relay extracts the system frame and sub-frame number, however these two values are insufficient to calculate the \textit{SchedulingRequest} parameter. In order to acquire this, the relay ignores this \textit{SR} message, which forces the victim to re-send another \textit{SR} message in the next period. After observing this second \textit{SR} message, the adversary can compute the periodicity $p$ and the \textit{subframe-offset} by simple subtraction. Finally, the \textit{sr-ConfigIndex} is obtained through a lookup operation in the 3GPP Table~10.1.5-1~\cite{3gpp-36213} where the \textit{sr-PUCCH-ResourceIndex} is the index of the radio resource used by the \textit{SR} message in the frequency domain.

At this stage, the relay adversary knows the \textit{schedulingRequestConfig} parameters and can use them to configure both its eNodeB and its UE interfaces. By dropping the first SR, however, the mobile-relay causes a time delay in the transmission of the \textit{RRCConnectionReconfigurationComplete} message. This time delay depends on the periodicity of SR, which normally is 10ms or 20ms. However, this delay will not trigger any connection failures given that (1) the guessing procedure is fast and only takes a maximum of two periods (e.g., 20ms) and (2) there are no timeouts available for receiving \textit{RRCConnectionReconfigurationComplete} messages by the eNodeB. Furthermore, this re-transmission procedure is a common occurrence which triggers failures only if the maximum number of re-transmissions is reached. The threshold, however, is sufficiently large (e.g., 64 re-transmissions for Carrier1) for our relay implementation to calculate the parameters without breaking the radio connection. We detail our procedure in Algorithm~\ref{guessing_scheduling_request_parameters}.

\vspace{5pt}
\noindent\textbf{Recovering \textit{CQI-ReportConfig} parameters.} This process is similar to the one used to recover \textit{schedulingRequestConfig} parameters, however it requires a few slight changes as follows. First, for Multiple Input Multiple Output (MIMO) connections the UE uses at least two antennas to send and receive radio signals. The 3GPP standard introduces the Rank Indicator (RI) parameter to measure to what extent the signals sent by one antenna interfere with the signals of the others, such that the eNodeB can adjust its transmission parameters and avoid serious interference. Therefore, the adversary needs to guess this \textit{ri-ConfigIndex} parameter only when using MIMO is detected. Second, when guessing \textit{schedulingRequestConfig}, the first \textit{SR} is dropped. However, when guessing \textit{CQI-ReportConfig}, the first message cannot be dropped since it affects the MCS used for downlink data which may not be correctly decoded if the \textit{CQI} message is dropped. However, processing the first \textit{CQI} message has no effect on the guessing procedure because the relay will receive a second message regardless of whether the first one is dropped or processed, as \textit{CQIs} are periodic messages.

\vspace{5pt}
\noindent\textbf{Recording VoLTE traffic.} Targeting VoLTE traffic specifically, for any reason, including recording, should not be possible when using EEA2 encryption algorithms which rely on non-deterministic encryption schemes such as AES-CTR. This however is not the case. By looking at the non-encrypted MAC sub-header at our mobile-relay, the attacker can learn the Logical Channel ID (LCID) of the sub-PDU (see Section~6 in \cite{3gpp-36321}). Because VoLTE traffic uses specific LCID 4 and LCID 5 it can be directly targeted by the adversary. In the following, we show how this recorded traffic is used to reveal information about a victim.

\subsection{VoLTE traffic analysis}
\label{call-signalling}

The main purpose of VoLTE traffic analysis is to process collected traffic and extract VoLTE activity logs, including signalling and voice logs. A related adversarial model to ours, which exploits protocol miss-implementations, has been used to recover encrypted voice data in LTE networks by Rupprecht et al.~\cite{rupprecht-call-2020}. Here we focus on recovering VoLTE logs using metadata traffic information protected by standard LTE/NR security, allowing our adversary to mount attacks against both LTE and 5G networks which correctly implement the standard mandated security features. As stated in Section~\ref{preliminaries}, VoLTE signalling is generated according to predefined templates and has static communication characteristics. Our work exploits these characteristics similarly to Xie et al.~\cite{8433136}, however, while they analyse plaintext Voice over WiFi (VoWiFi) traffic collected on a malicious Access Point (AP), we deal with the more complex case of extracting meaningful logs from intercepted LTE/5G traffic, which uses both IPsec and standard EEA2 user-plane encryption.

\vspace{5pt}
\noindent\textbf{IP packet reassembly.} Mobile LTE/5G networks use fragmentation to efficiently transfer oversized application messages (e.g., VoLTE, Hypertext Transfer Protocol (HTTP)). When transmitting data over a mobile connection, each TCP (or UDP) segment is first encapsulated in an IP packet and then in a PDCP layer packet. Each PDCP packet contains a \textit{Sequence Number} and an encrypted and integrity protected IP packet as payload. Segmentation or concatenation can happen at lower layers if required by the protocol, but because encryption only happens at the PDCP layer, an adversary can revert these operations and restore PDCP packets. A passive mobile-relay adversary can further obtain information about the direction $dir$ (i.e. uplink or downlink) and arrival time $time$ of PDCP packets by simply observing traffic.

The adversary, however, does not have any information about the contents of PDCP packets. In order to make sense of these and reconstruct meaningful VoLTE messages that can be analysed we leverage generic knowledge about network protocols. First, we assume that each TCP or (UDP) segment is efficiently used according to the Maximum Transmission Unit (MTU), i.e. the size of all fragments in a sequence except the last one is equal to the MTU at the moment of segmentation. The MTU is determined from the Maximu\_SDU\_size contained in NAS messages and is same as the one observed by the attacker's UE. Using this assumption, we give an efficient packet reassembly algorithm. Briefly, based on observation, VoLTE related packets are usually split into three fragments. Our algorithm tries to reconstruct these sequences by looking at neighbouring packets and trying to allocate them to a category, e.g., first, middle, or last, based on the relationship between their real size and their MTU. Once reassembled, the adversary requires some protocol context relevant info to the type of VoLTE traffic (i.e. TCP, UDP, TCP over IPsec, or UDP over IPsec) to calculate the size of the SIP signalling payload by subtracting all protocol headers from IP packet length. We obtain this information from Control Information (CI) packets (i.e. SYNC, FIN, ACK) which are transferred between peers when TCP connection setup, tear down, or maintenance. Although CI packets are encrypted, the adversary is still able to locate them by examining packet size, e.g., the TCP header length of SYNC, SYNC\_ACK, and ACK are 40, 32, and 20, respectively.

\vspace{5pt}
\noindent\textbf{VoLTE signalling identification.}
After IP packets have been reassembled from encrypted PDCP traffic, the adversary needs to identify VoLTE data streams. The main challenge is to link the encrypted messages to specific VoLTE operations such as \textit{Invite}, \textit{Cancel}, and restore the communication logs. This can be accomplished as follows. First, a one-off operation is required, where the adversary builds a database which encodes VoLTE message characteristics corresponding to each type of operation. This process can be accomplished easily by using standard diagnostic tools, e.g., SCAT~\cite{hong2018peeking}, to analyse network traffic on an attacker controlled UE. While this traffic is usually encrypted at the IPSec level, all the session keys can be obtained with readily available tools such as SIMTrace~\cite{simtrace}. With the decrypted VoLTE messages, the adversary is able to construct a message characteristics database specific to a victim network carrier such as the one shown in Table~\ref{tab:signal_len_map}. Using this database the adversary is able to map encrypted VoLTE messages to their corresponding operations by evaluating their direction, encrypted size and type of operation. We observe that message characteristics depend on the VoLTE software provisioned in the baseband firmware, and the carrier used, are consistent for same model devices, and are fairly static between models.

At the end of the mapping operation, the adversary is able to extract complete VoLTE signalling logs which contain the following five features: (1) \textit{identity}: the victim's identity such as Subscriber Concealed identifier (SUCI), IMSI, phone number; (2) \textit{timestamp}: the time of day of the VoLTE call; (3) \textit{call direction}: incoming or outgoing call for victim; (4) \textit{establish status}: the response of callee (i.e. accepted, declined or missed); (5) \textit{termination cause}: which UE ended the call session and for what reason (e.g., caller cancelled during ring period, callee hang-up during conversation); (5) \textit{call duration}: the duration time (in second) of this VoLTE call.

\vspace{5pt}
\noindent\textbf{VoLTE voice activity.} In addition to the features mentioned above, the adversary is also able to extract the victim's voice activity to an accuracy window of 20ms by analysing \textit{Comfort Noise} frames.

To do this, first, the adversary refines voice related traffic by filtering out RTCP packets from the collected DRB3 traffic because RTCP packets can be transferred on the DRB3 or the DRB2 alongside RTP which depends on the carrier's configuration. RTCP packets can be easily identified based on their fixed size (e.g., 128 or 140 bytes). The \textit{Comfort Noise} frames are encoded within RTP packets as the special frames which contain background noise parameters instead of encoded audio data, and they are generated only when Voice Activity Detection (VAD) detects that the speaker has not spoken in the last sample period. Given that no actual data needs to be encoded in these frames, the size of \textit{Comfort Noise} frame is 6 bytes which is smaller than others (e.g., Adaptive Multi-Rate Wideband (AMR-WR) generates 132 or 477 bits)~\cite{3gpp-26201,3gpp-26090}. Additionally, \textit{Comfort Noise} frames have a lower re-transmission frequency, as low as one packet every 160 ms whereas other frames are re-transmitted every 20 ms~\cite{3gpp-26102,3gpp-26201}. Once a \textit{Comfort Noise} frame is observed, the adversary automatically learns that the victim has not spoken in the last 160 ms.

\begin{figure}[ht]
    \centering
    \includegraphics[width=0.9\columnwidth]{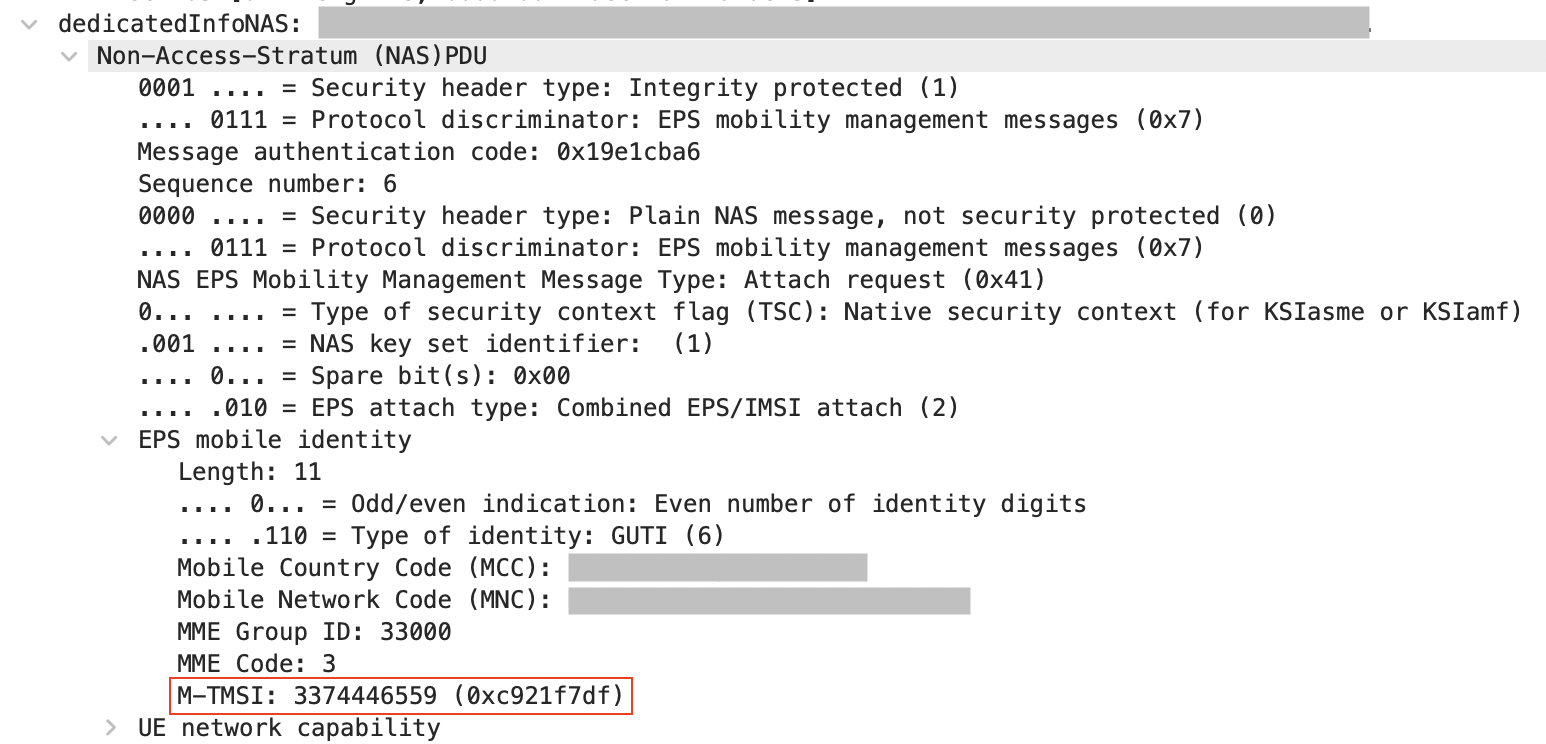}
    \caption{An example of \textit{Attach Request} which uses GUTI as a user identifier. The adversary modifies the \textit{M-TMSI} to 0x12345678 in order to break the security context established by the previous AKA procedure to force the network to reinitialize the authentication with the UE.}
    \label{fig:attach_request}
\end{figure}

\subsection{Identity mapping using VoLTE}\label{link-imsi}
The main goal of identity mapping is to link the collected network identifier (i.e. IMSI, SUCI, Globally Unique Temporary Identifier (GUTI)) to the victim's real-word identity (i.e. phone number) to further monitor a specific victim's VoLTE activities. First, we discuss our \textit{passive mapping with call capability} which maps anonymised identity (i.e. SUCI and GUTI) to the real-world identity. To this end, the adversary needs to make a VoLTE call towards the victim to trigger VoLTE traffic between the victim's UE and the IMS. Then, the collected traffic is analysed to obtain the victim's VoLTE logs (Section~\ref{call-signalling}). The analysed traffic is combined with details related to the call, available to the attacker from its own UE, in order to link the phone number of the victim to its identity. This procedure does not require the victim to perform any response action related to the incoming call, because several signalling messages (e.g., Invite, Ring) are exchanged between the victim UE and the IP Multimedia Subsystem (IMS) before the actual ringing event on the UE happens. Observing these messages in the logs is sufficient to perform the correlation.

This is mostly a one-off operation because even temporary identities remain the same for extended periods of time~\cite{hong2018guti,shaik-practical-2015}. This is also supported by our observation of GUTI reallocation, which is discussed in Section~\ref{exp: analysing VoLTE signalling log}. When the victim's UE connects to our mobile-relay again, there is no need to repeat this mapping procedure if the victim's GUTI has not changed since the previously observed value.

The stronger \textit{active mapping} procedure needs an additional step in order to break the Evolved Packet-switched System (EPS) security context. This procedure is similar to the Uplink IMSI Extractor proposed by Erni et al.~\cite{arxiv.2106.05039}, which overshadows the uplink \textit{Attach/Service Request} message. However, our attack remains undetectable because we do not trigger a \textit{Security Mode Reject} fault at victim UE.

In Fig.~\ref{fig:attach_request}, we show an example of \textit{Attach Request} message containing user's GUTI. We modify the M-Temporary Mobile Subscriber Identity (M-TMSI) value in this message to 0x12345678 using our mobile-relay and keep the remaining values unchanged. This causes the message authentication code of this message to become invalid, which in turn, causes the carrier to respond with an \textit{Identity Request} message which forces the UE to start the Authentication and Key Agreement (AKA) procedure~\cite{3gpp-24301}. The adversary is now able to obtain the victim's IMSI from the subsequent plaintext \textit{Identity Response}. The mapping procedure remains the same as the previous \textit{passive mapping}.

\section{Real-world results}\label{exp: real-world results}
We verify the feasibility of our attack using four COTS UEs which we connect to two commercial carriers. In the following, we describe our experimental setup and continue with our test procedures and results.

\subsection{Experimental setup}
In Fig.~\ref{fig:testbed} we present our experimental setup, and we depict these components and their functions as follows:

\begin{figure}[t]
    \centering
    \includegraphics[page=1,width=0.9\columnwidth]{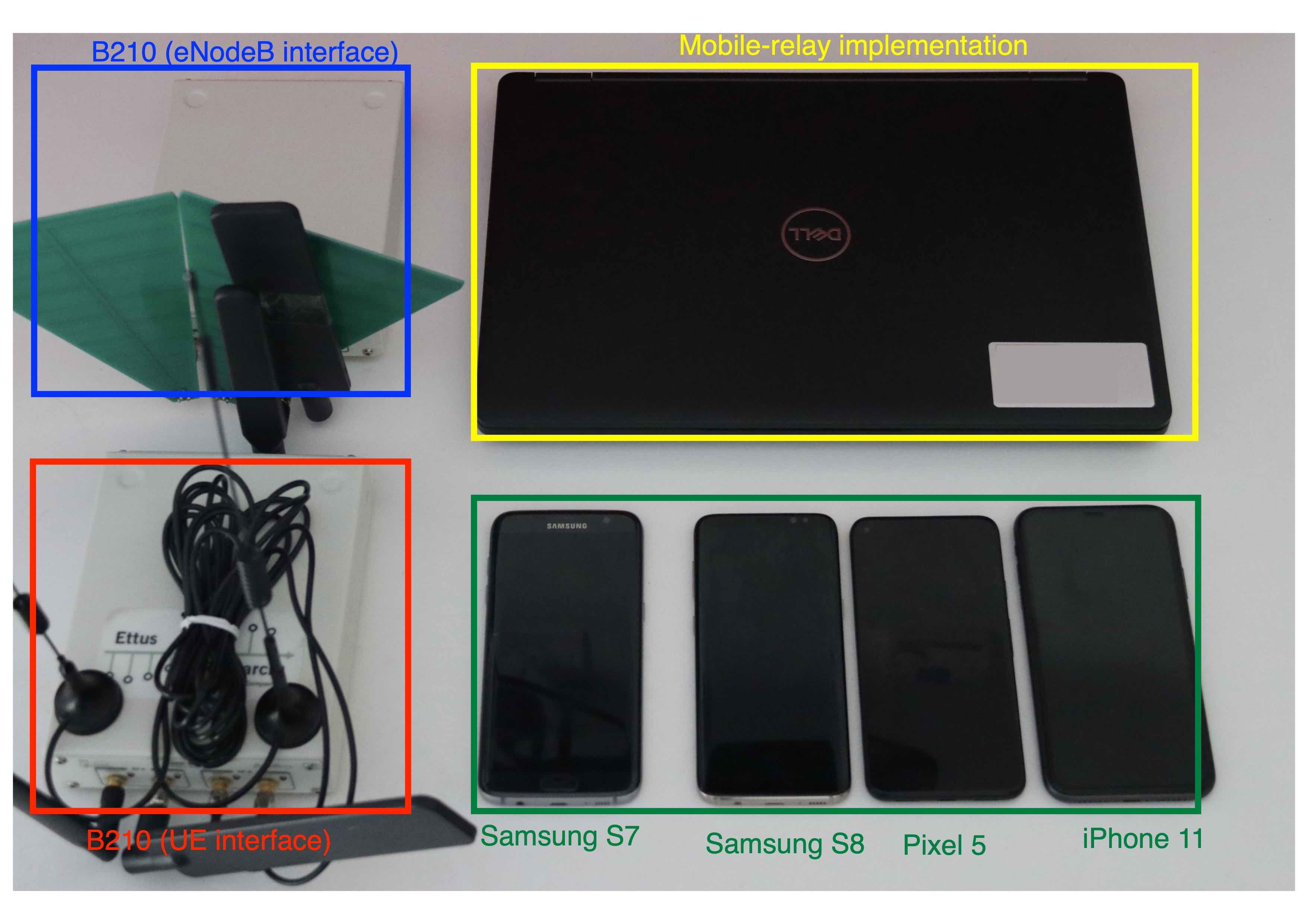}
    \caption{Experimental setup. Our mobile-relay software implementation runs on the laptop computer. Two USRP B210 SDRs are connected, one acting as an eNodeB and the other as a UE interface.}
    \label{fig:testbed}
\end{figure}

\begin{itemize}
    \item \textbf{UEs.} We use Android Debug Bridge (ADB) to operate Android phones, e.g., toggling airplane mode and dialling VoLTE calls. Samsung S7 and S8 allow us to collect Control Plane (CP) and User Plane (UP) information from the diagnostic interface using SCAT~\cite{hong2018peeking}. For iPhone 11, we toggle airplane mode using the \textit{Mirror iPhone} via Apple Watch and capture UP traffic using rvictl~\cite{rvi}. The OS, chipset and baseband versions of the tested UEs are shown in Table~\ref{tab:overview_of_test_phones}.
    \item \textbf{Mobile-relay.} Our mobile-relay runs on Arch Linux with Kernel 5.17.1-arch1-1 and Intel i5-8250U, and consists of two Ettus USRP B210 controlled by a modified version of the srsRAN v21.10~\cite{srsran} software stack. One B210 acts as the eNodeB interface towards the victim UE(s), while the other simulates a UE interface towards the commercial eNodeB. The eNodeB component copies the configuration from the targeted commercial eNodeB.
    \item \textbf{Commercial eNodeB and carriers.} We connect our mobile-relay to the commercial eNodeB and use specific commercial network USIM cards on the victim UE to mimic real-world use. We test our attacks on two major commercial network carriers: Carrier1 and Carrier2. Carrier1 uses MIMO while Carrier2 uses Carrier Aggregation (CA).
\end{itemize}

\begin{table}[t]
    \centering
    \begin{tabular}{l|l|c|c}
        \hline
        \multicolumn{2}{l|}{Parameters} & Carrier1 & Carrier2 \\ \hline
        \multirow{5}{*}{CQI} & \textit{cqi-PUCCH-ResourceIndex} & \cmark & \textdagger \\
        & \textit{cqi-pmi-ConfigIndex} &  \xmark & \xmark \\
        & \textit{cqi-FormatIndicatorPeriodic} & \cmark & \cmark \\
        & \textit{ri-ConfigIndex} & \cmark & \textdagger \\ \hline
        & \textit{simuaneousAckNackAndCQI} & \cmark & \cmark \\
        \multirow{3}{*}{SR} & \textit{sr-PUCCH-ResourceIndex} & \textdagger & \textdagger \\
        & \textit{sr-ConfigIndex} & \xmark & \xmark \\
        & \textit{dsr-TransMax} & \cmark & \cmark \\
        \hline
    \end{tabular}
    \vspace{5pt}
    \caption{Physical layer configuration parameters as observed for Carrier1 and Carrier2 where \cmark ~represents static values, \textdagger ~a small search space and \xmark ~that no optimisations are possible.}
    \label{tab:physical_config_for_providers}
\end{table}

\begin{table*}[tb]
    \centering
    \begin{tabular}{|l|l|l|l|l|l|l|l|l}
        \hline
         \multirow{2}{*}{Phone} &  \multirow{2}{*}{OS Ver.} & \multirow{2}{*}{Chipset} & \multirow{2}{*}{Baseband Ver.} & \multicolumn{2}{c|}{Carrier1} & \multicolumn{2}{c|}{Carrier2} \\
         \cline{5-8}
         & & & & AKA & Bearers & AKA & Bearers \\ \hline
         iPhone 11 & 15.4.1 & Apple A13 & 3.02.01 & \cmark & \cmark & \cmark & \xmark \\ \hline
         Samsung S7 & 8.0.0 & Qualcomm & G935FXXU8EUE1 & \cmark & \cmark & \cmark & \cmark \\ \hline
         Samsung S8 & 9.0 & Exynos & G9500ZHS6DUD1 & \cmark & \cmark & \cmark & \xmark \\ \hline
         Pixel 5 & 12.0 & Qualcomm & g7250-00188-220211-B-8174514 & \cmark & \cmark & \cmark & \xmark\\
         \hline
    \end{tabular}
    \vspace{5pt}
    \caption{Overview of the configurations of UEs and network carriers where \cmark~ means that the UE has complete functionality with the carrier and \xmark~ that the UE only has partial functionality due to hardware limitations of B210 SDR. Carrier1 requires use of MIMO. For this carrier, all four phones successfully complete AKA authentication procedure and successfully set up bearers (e.g., Internet, VoLTE). Carrier2 requires use of Carrier Aggregation. With this carrier tested phones complete the AKA procedure but only the Samsung S7 is able to set up EPS bearers. This is because Carrier Aggregation (CA) is not feasible when using B210 SDRs.  }
    \label{tab:overview_of_test_phones}
\end{table*}

\subsection{Experimental procedure}
In the following, we give a high-level description of our experimental procedures. After, we continue with details and specific insights learned from our tests.

\begin{enumerate}[1.]
    \item \textbf{Monitoring the victim UE.} We first activate the airplane mode on victim UE. After starting mobile-relay, we disable airplane mode and wait for victim UE to connect to our relay. Once the UE is registered to the network, we perform a number of VoLTE activities, such as dialling, answering and declining calls, in order to generate VoLTE traffic. We continuously monitor control plane traffic at the relay level. We immediately start the guessing procedure when \textit{RRCConnectionReconfiguration} message is observed.
    \item \textbf{Collecting identities.} For the \textit{passive attack}, we collect victim's identities that are contained in \textit{Attach/Service Request} messages. For the \textit{active attack}, we modify the \textit{Attach/Service Request} message which triggers a break in the EPS security context between the victim UE and the network, due to integrity protection checks failing. This forces the victim to identify itself using long term IMSI identity.
    \item \textbf{Analysis of VoLTE logs.} We use the method described in Section~\ref{call-signalling} to extract the victim's VoLTE activities, including signalling logs and voice logs.
    \item \textbf{Identity mapping.} In order to map the collected identity to an actual phone number, we make a VoLTE call towards the victim UE from the attacker controlled UE. By analysing the corresponding VoLTE traffic between the victim and the attacker, we can identify which phone is associated with the dialled phone number.
\end{enumerate}

\subsection{Guessing physical layer parameters}
\label{exp: guessing physical layer parameters}

As introduced in Section~\ref{volte-related-packets}, the adversary needs to know physical layer parameters in order for the mobile-relay to maintain the radio connections. We develop a \textit{guessing} procedure for these, which requires the adversary to observe the parameter patterns of the radio bearers contained in the \textit{RRCConnectionReconfiguration} messages.

\vspace{5pt}
\noindent\textbf{Physical parameters' analysis procedure.} We collect the Control Plane (CP) data for 60 hours for each carrier. Collected data shows that most parameters of \textit{physicalConfigDedicated} are fixed while only \textit{cqi-ReportPeriodic} and \textit{schedulingRequestConfig} have slight variations. We summarise the major parameters in Table~\ref{tab:physical_config_for_providers}. Parameters \textit{cqi-FormatIndicatorPeriodic}, \textit{simuaneousAckNackAndCQI} and \textit{dsr-TransMax} always have the same values, while \textit{cqi-pmi-ConfigIndex} and \textit{sr-ConfigIndex} refreshed every time. For Carrier1, we observed that parameters \textit{cqi-PUCCH-ResourceIndex} and \textit{ri-ConfigIndex} are fixed. These however vary between a small set of values for Carrier2. The \textit{sr-PUCCH-ResourceIndex} parameter has several values both for Carrier1 and Carrier2.

By observing this pattern we were able to reduce the complexity of guessing real-word parameters as follows: (1) for fixed parameters, we just set them to the observed value every time; (2) for changing parameters, which have limited options, we first analyse their occurrence frequency and then try the options in priority decreasing order. For example, \textit{sr-PUCCH-ResourceIndex} for the Carrier2 has 28 options, however the top option takes 53.14\% and top-five options take 83\%. Finally, (3) we find that the periodicity of \textit{SR} are fixed for each LCID in both Carrier1 and Carrier2 (e.g., Carrier2 sets periodicity as 20, 10, 10 for LCID 5, 6 and 7, respectively). This stable periodicity provides the ability to immediately calculate \textit{sr-ConfigIndex} after the first request has arrived (as shown in Line 7-8 in the Algorithm~\ref{guessing_scheduling_request_parameters}).

\begin{figure}
    \centering
    \includegraphics[width=0.9\columnwidth]{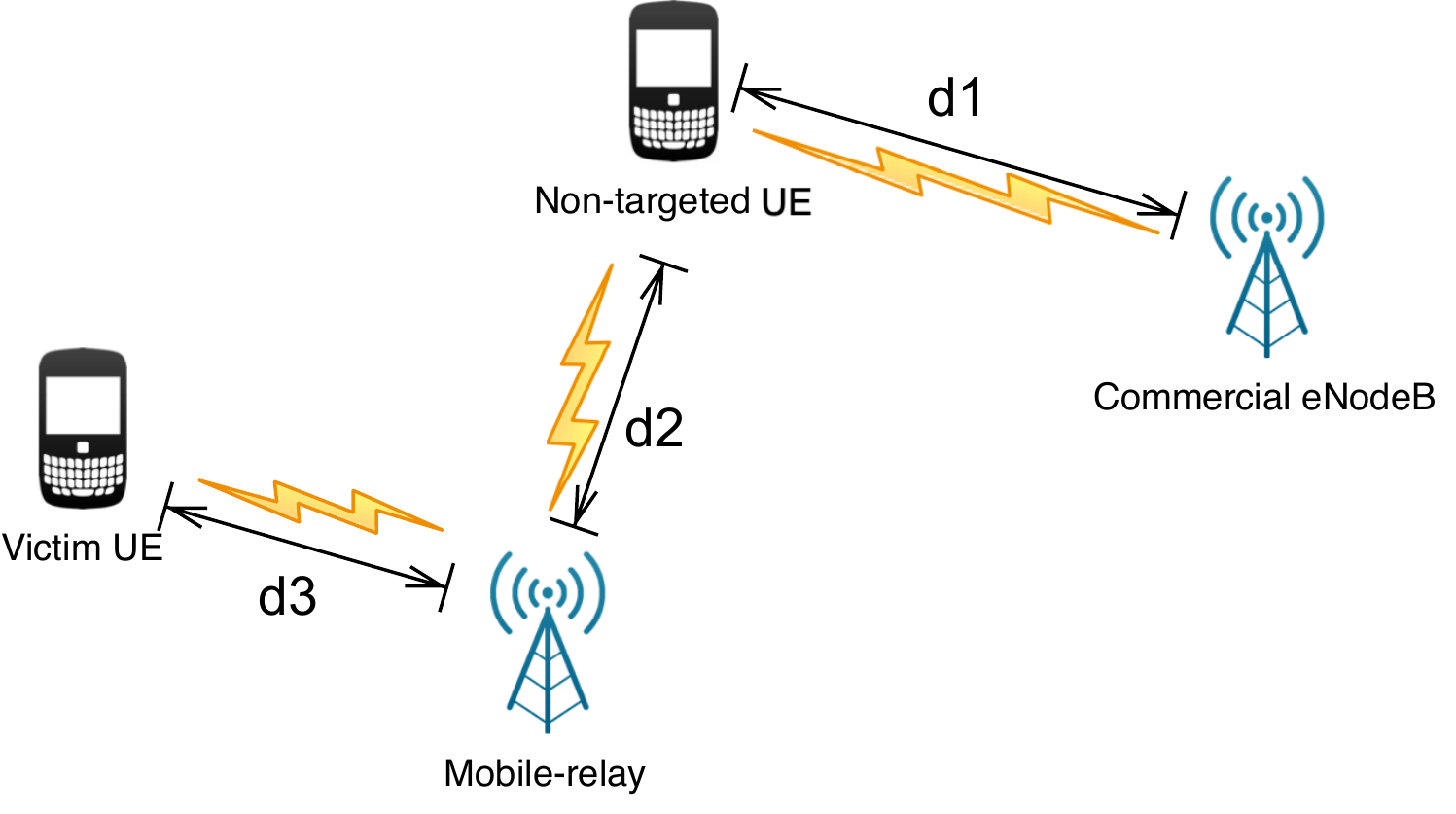}
    \caption{Parameter detection using radio signal interference. Non-targeted UE connects to commercial eNodeB with distance $d1$ and targeted UE connects to mobile-relay with distance $d3$. The distance between non-targeted UE and mobile-relay is $d2$. Since $d2$ is not equal to $d1$, the propagation delays of these two parts are different.}
    \label{fig: air-interference}
\end{figure}

\vspace{5pt}
\noindent\textbf{Dealing with radio signal interference.} During the guessing period, a major challenge is dealing with radio signal interference as the mobile-relay opens all proper resources in frequency and time domain to look for specific victim UE's Physical Uplink Control Channel (PUCCH) messages (\textit{SR} and CQI). Messages transmitted from non-targeted UEs can be received by the mobile-relay which causes interference in distinguishing between messages originating from victim UE and the ones from the non-targeted UEs. Fig.~\ref{fig: air-interference} shows such an environment observed in a real-world relay deployment, where a victim UE and a non-targeted UE connect to the mobile-relay and to the commercial eNodeB, separately. The mobile-relay not only receives the radio signals transmitted from the victim UE but also from the non-targeted UE.

However, using distance measurements the adversary can distinguish between a victim UE connected to the relay and non-targeted UEs as follows. Assuming the setup in Fig.~\ref{fig: air-interference}, in the normal case, the distance $d1$ between a non-targeted UE and commercial eNodeB is different from the distance $d2$ between the same non-targeted UE and the mobile-relay, therefore, one can compute the propagation delay of both paths as $d1/c$ and $d2/c$ respectively. eNodeB measures this propagation delay also and uses the \textit{Time Advance} (\textit{TA}) parameter to instruct UEs to align their internal clocks by adjusting uplink data transmission time to be slightly ahead i.e. $2*d1/c$ (see Section 8 in \cite{3gpp-36211} and Section 4.2.3 in \cite{3gpp-36213}). Since the non-targeted UE are aligned to the commercial eNodeB rather than the mobile-relay, the time delay of the received PUCCH messages transmitted from non-targeted UE's at mobile-relay is $(d2-d1)/c$. However, the time delay of victim UE's messages at mobile-relay is $0$ since victim UE has aligned to mobile-relay using \textit{TA}. Another signal feature which can be leveraged to identify the victim UE is the Signal-to-Noise Ratio (SNR) which indicates the quality of the radio channel used by this received message. The higher the SNR, the better the signal quality. In this work, we use these two features (i.e. TA and SNR) of the radio channel to determine if the received messages are transmitted by victim UE or not.

\begin{figure}
    \centering
    \includegraphics[width=0.9\columnwidth]{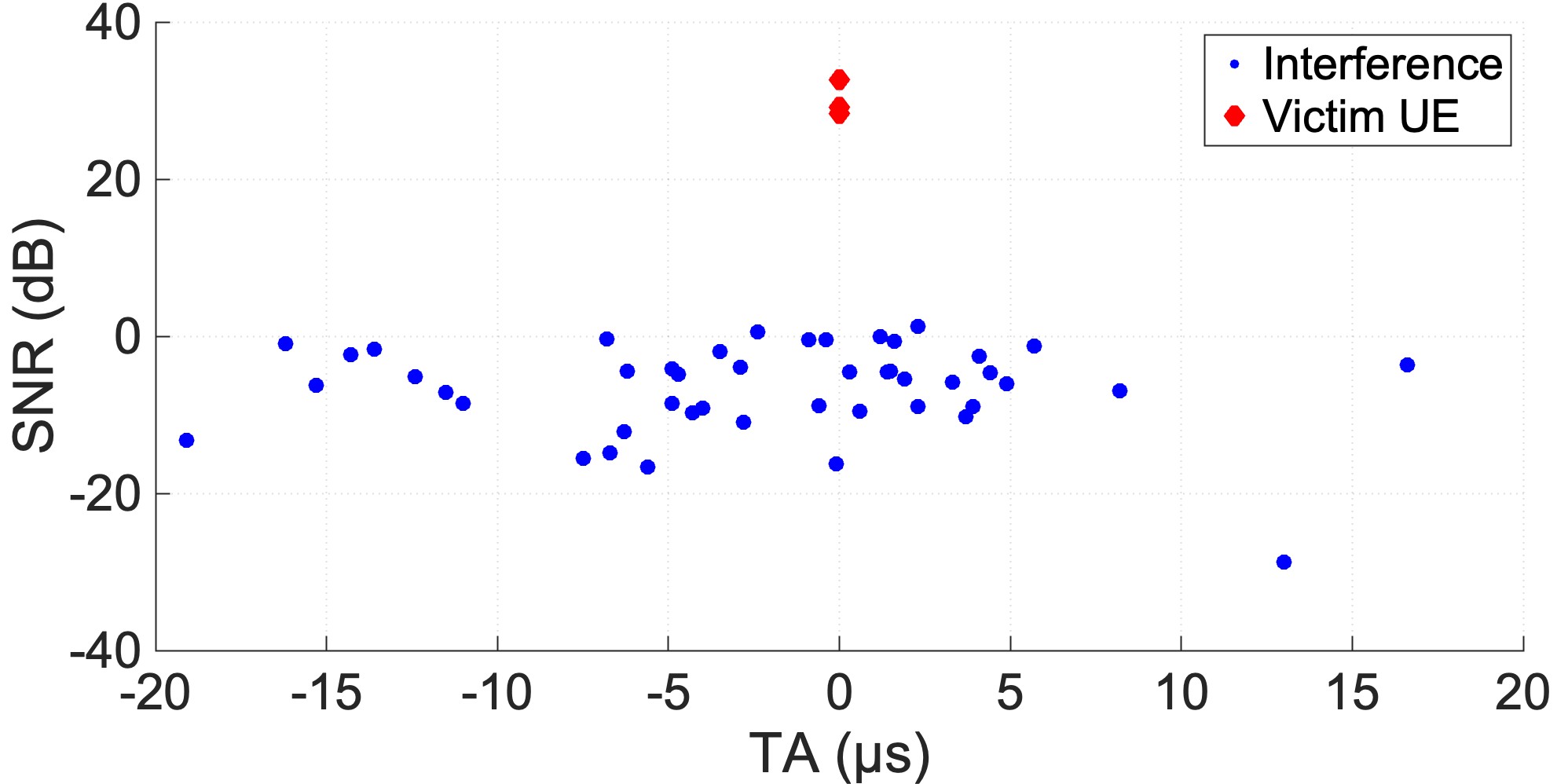}
    \caption{The scatter of \textit{TA} and \textit{SNR} of the messages received by mobile-relay during a \textit{guessing} period. The messages transmitted from the victim UE have higher \textit{SNR} above $20$dB and stable \textit{TA} as $0\mu$s, while the \textit{SNR} for other messages transmitted from non-targeted UE is quite low and \textit{TA} of these messages are distributed between $-20\mu$s to $20\mu$s.}
    \label{fig: radio signal features during guessing}
\end{figure}

In Fig.~\ref{fig: radio signal features during guessing}, we show real-world measurements for \textit{TA} and \textit{SNR} as obtained from intercepted PUCCH messages during a \textit{guessing} period. As expected, the \textit{TA} of victim UE's messages are located around 0$\mu$s while those from others are distributed between $-20\mu$s to $20\mu$s. The \textit{SNR} of victim UE's messages are quite high, above $20$dB, in contrast, the \textit{SNR} of others is quite lower, almost all of them below $0$dB. Based on these observations, our relay is able to accurately identify the targeted UE and adjust the physical parameters accordingly.

\vspace{5pt}
\noindent\textbf{Connectivity results.} All evaluated UEs are able to complete the authentication procedure, and setup default Internet, VoLTE signalling and voice bearers as shown in Table~\ref{tab:overview_of_test_phones}. Complete VoLTE functionality is achieved for Carrier1. For Carrier2, however, bearers are successfully established only for the Samsung S7. This is caused by hardware limitations of USRP B210, specifically by the Carrier Aggregation (CA) which requires at least two channels running at different carrier frequencies. Unfortunately, the B210 only supports one. In the case of the S7, the baseband firmware first establishes one connection to the eNB and then attempts a secondary one. This, however, is unsuccessful when using the B210 due to the above mentioned limitations. Unlike other firmware though, the S7 does not disconnect the first established connection upon the failure of the second.

In order to evaluate the success rate of guessing physical layer parameters, we execute the connection procedure between the victim UE and the mobile-relay 60 times. Our results show a success rate of 91.67\%. When investigating the root causes for the occasional failures, we observe that most are caused by hardware limitations related to the attacker processing power. Effectively, our implemented attacker is unable to process data at the required rates such that it can decode all candidate resource blocks and identify the targeted scheduling requests. We estimate that attackers with better hardware (e.g., faster CPUs) will easily achieve better results.

\subsection{Analysing VoLTE signalling log}\label{exp: analysing VoLTE signalling log}
The analysis of the communication characteristics of VoLTE signalling is an important step before moving on to real-world experiments. Here, we simulate four common scenarios to generate and analyse VoLTE traffic and evaluate traffic identification performance. These scenarios, and the specific SIP messages encountered, are briefly described in the following.

\begin{enumerate}[(1)]
\item \textit{Call cancelled during ringing by the caller.} In this scenario, the caller sends an \textit{Invite} message to the callee to trigger the new call session setup. The callee responds with a \textit{Ring} message to the caller. Upon receiving this message, the caller terminates this session by sending its own \textit{Cancel} message to the callee.

\item \textit{Call cancelled during conversation by the caller.} This is similar to the previous scenario with the main difference is the call session is cancelled during conversation by the caller. After the callee responds with \textit{Ring} the caller does nothing and waits for the \textit{OK (Invite)} response which is sent by the callee when the incoming call is accepted. Then, after the conversation starts and audio data is observed on DRB3, the caller terminates the call by sending a \textit{Bye} request message.

\item \textit{Call declined by the callee.}  In this scenario, the callee responds with a \textit{Busy Here} message after \textit{Ring} message to terminate the session between itself and the IMS. After the IMS receives the \textit{Busy Here} response, it redirects the call session to the callee's voice mail if voice mail is enabled, otherwise, IMS sends \textit{Busy Here} response to the caller to terminate the session between the caller and IMS.

\item \textit{Call cancelled during conversation by the callee.} This is similar to the second scenario with the difference being that the \textit{Bye} request message is sent from the callee rather than the caller.
\end{enumerate}

\vspace{5pt}
\noindent\textbf{VoLTE signalling analysis procedure.} We execute the scenarios above on a Samsung S7, a Samsung S8 and an iPhone with Carrier1. We also test the iPhone with Carrier2 where we collect and analyse VoLTE signals. Our test scenario involves making a VoLTE call between two victim UEs, one connected through our mobile-relay and the other connected directly to the network carrier. We repeat each scenario five times and collect 1386 SIP messages in total. Even though the calls are identical, during our tests, we observe that the number of generated SIP messages is not constant for each call as shown in Table~\ref{tab:signal_len_map}. For example, the Samsung S7 sends a \textit{200 OK (Update)} message, however, the S8 and iPhone 11 do not. The collected data additionally shows that (1) the IPsec configurations for carriers 1 and 2 are the same, with one exception, Carrier2 encrypts IPsec payloads using \textit{AES-CBC} while Carrier1 uses plaintexts; (2) SIP messages can be sent with either \textit{TCP-over-IPsec} or \textit{UDP-over-IPsec}; (3) the MTUs are 1308 and 1276 for uplink and downlink for Carrier2, and 1212 for both uplink and downlink for Carrier1. We further analyse the size of each SIP message and find the communication characteristics as shown in Table~\ref{tab:signal_len_map}. We detail these in the following. 

\begin{enumerate}[(1)]
    \item For most SIP messages the size is relatively constant, showing only minor variations, while the size falls within two or three byte ranges for some messages (e.g., downlink \textit{183 Session Process} message). Falling into different byte ranges is determined to be caused by they are generated in different contexts though they share the same operation type. For example, a caller receives a \textit{200 OK (Invite)} response message in both the callee accepted and declined scenarios, however, the former establishes the normal conversation and the latter redirects the call to the callee's voice mail.
    
    \item For downlink SIP messages, the signal size is similar within a carrier even though the UEs are different. For example, within Carrier1, the size of downlink \textit{Invite} message for tested iPhone11, Samsung S7 and S8 are similar as $2371\pm6$, $2358\pm8$ and $2357\pm5$. This is reasonable because downlink signals are generated by the carrier's IMS which keeps the same. However, for different carriers, the downlink size is various since the carriers' IMSs are different. The downlink \textit{Invite} messages of iPhone 11 have different lengths i.e. for Carrier2 messages are located in the $[2219\pm2,2000\pm0]$ bytes range while for Carrier1 they are usually of constant length e.g., $2371\pm6$ bytes.
    
    \item For uplink SIP messages, the signal size is related to carrier and phone brand. The uplink characteristics are similar for the same phone brand within a carrier. For example, the size of uplink \textit{Invite}, \textit{100 Trying (Invite)}, \textit{183 Session Process} messages for Samsung S7 and S8 are similarly as $2479\pm0$, $338\pm1$, $1437$ and $2494$, $336$, $1435$ bytes.
\end{enumerate}

\begin{figure}
    \centering
    \includegraphics[width=0.9\columnwidth]{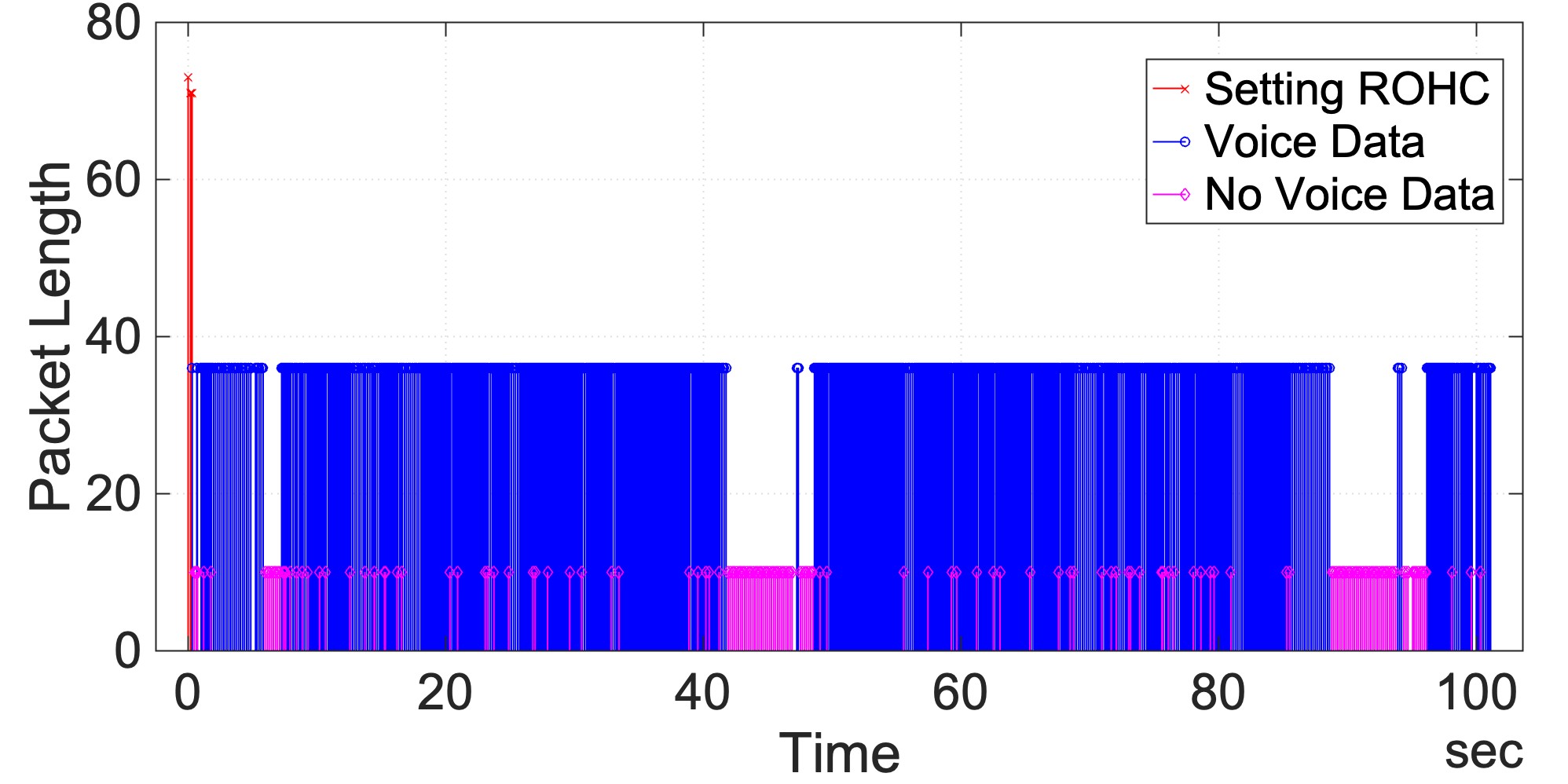}
    \caption{Time-sorted downlink RTP traffic representation. The sizes of the frames which contain audio data (blue) are significantly larger when compared to \textit{Comfort Noise} frames (purple). The first several frames (red) are much larger than the rest because the Robust Header Compression (ROHC) context has not been established.}
    \label{fig:conversation_log}
\end{figure}

\vspace{5pt}
\noindent\textbf{Real-word results.} We make 16 VoLTE calls on the Samsung S7 and S8 with Carrier1 to evaluate our attack. We set the MTUs as the observed value as 1212 bytes for both uplink and downlink, and we use the method introduced in Section~\ref{call-signalling} to preprocess collected encrypted PDCP packets and identify encrypted SIP messages using databases (as shown in Table~\ref{tab:signal_len_map}). We record 130 SIP messages with our relay and we map them to specific VoLTE operations with 83.07\% accuracy. We further analyse the causes where we fail to correctly identify messages and find that most are caused by the size similarities between operations, e.g., the size of uplink \textit{180 Ring} message from the Samsung S7 with Carrier1 is $877\pm1$ bytes while \textit{486 Busy Here} message has $878\pm1$ bytes. Therefore, we further revise the signalling log based on context (e.g., \textit{486 Busy Here} response can not happen before \textit{180 Ring (Invite)} response), which enables us to achieve 100\% accuracy. Fig.~\ref{fig:sip_log_analysis_result} shows an example of the recovered SIP messages from a victim UE.

\subsection{Monitoring voice activity}
\label{exp: monitoring voice activity}

In order to evaluate voice activity, we set up a VoLTE call from the iPhone 11 to a victim which uses Samsung S7 UE. Once the call is established, an audio sample is played from the iPhone 11. We terminate the call after 105 seconds. The call generates 3353 RTP packets in the downlink direction and 4864 packets in the uplink. In order to identify RTP packets which contain \textit{Comfort Noise} frame, we set a threshold at 10 bytes per message (6 bytes for \textit{Comfort Noise} frame, 1 byte for AMR header and 3 bytes for Robust Header Compression header). We show the analysis result of downlink RTP packets in Fig.~\ref{fig:conversation_log}. We can see that the downlink traffic has a bigger bit-rate when the callee is speaking than during silence periods. The large packet size observed at the start of the conversation is caused by the ROHC context which has not been established. The complete voice activity is obtained by analysing both uplink and downlink traffic.

\subsection{Mapping victims' identity}
\label{exp: linking identity}

In the following, we present the results of Globally Unique Temporary Identifier (GUTI) reallocation observed with Carrier1 and Carrier2, followed by the evaluation of \textit{passive mapping with call capability} and \textit{active mapping}.

We connect the Samsung S7 and the S8 to Carrier1 and Carrier2 for 60 hours and make calls every 10 minutes to collect Control Plane (CP) data. We find that the GUTI remains constant during the whole observed period. Therefore, the mapping between the victim's GUTI and the phone number is valid for extended periods of time and the VoLTE calls towards the victim are not frequently required.

\begin{figure}
    \centering
    \begin{subfigure}[b]{\columnwidth}
        \centering
        \includegraphics[width=0.9\columnwidth]{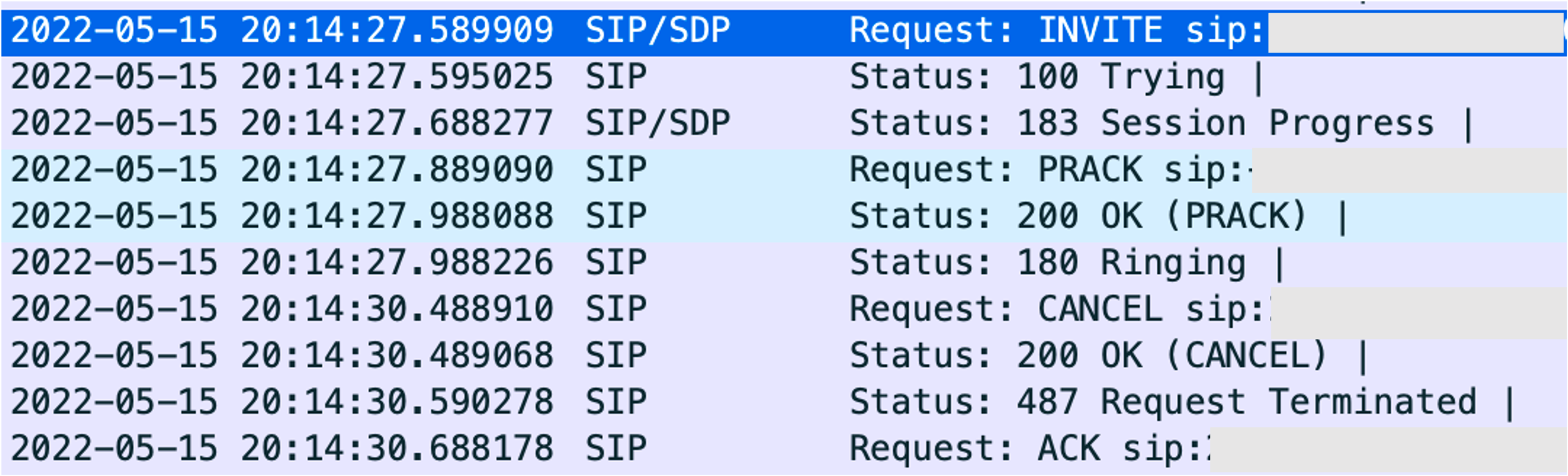}
        \caption{Reference VoLTE log as observed on the victim UE.}
        \label{fig:sip_log}
    \end{subfigure}
    
    \vspace{5pt}
    
    \begin{subfigure}[b]{\columnwidth}
        \centering
        \includegraphics[width=0.9\columnwidth]{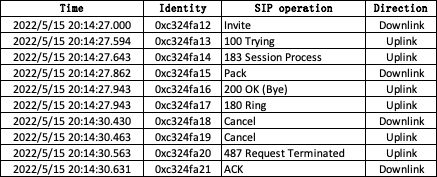}
        \caption{VoLTE log as observed by the mobile-relay adversary.}
        \label{fig:sip_log_analysis_result}
    \end{subfigure}
    \caption{VoLTE signalling logs from both the victim's UE and the mobile-relay adversary. The log recovered by the mobile-relay adversary is identical to the reference log. This can be used by an adversary to link the victim's identity to phone number.}
    \label{fig:volte_signalling_log}
\end{figure}

In Fig.~\ref{fig:volte_signalling_log} we show the results of \textit{passive mapping}. The real signalling log is shown in Fig.~\ref{fig:sip_log} and the VoLTE signalling analysis results obtained at our mobile-relay are shown in Fig.~\ref{fig:sip_log_analysis_result}. By using the sequence between the messages and their timestamps, an attacker can easily associate a known phone number with the observed activity. And in the case of an \textit{active mapping} attack, the victim's UE is forced to register to the network through a new Authentication and Key Agreement (AKA) procedure, which further reveals the victim's long term IMSI identity.

\section{Relay evaluation in 5G networks}
\label{5G-eval}

\begin{figure}
    \centering
    \begin{subfigure}[t]{0.43\columnwidth}
        \centering
        \includegraphics[height=1in]{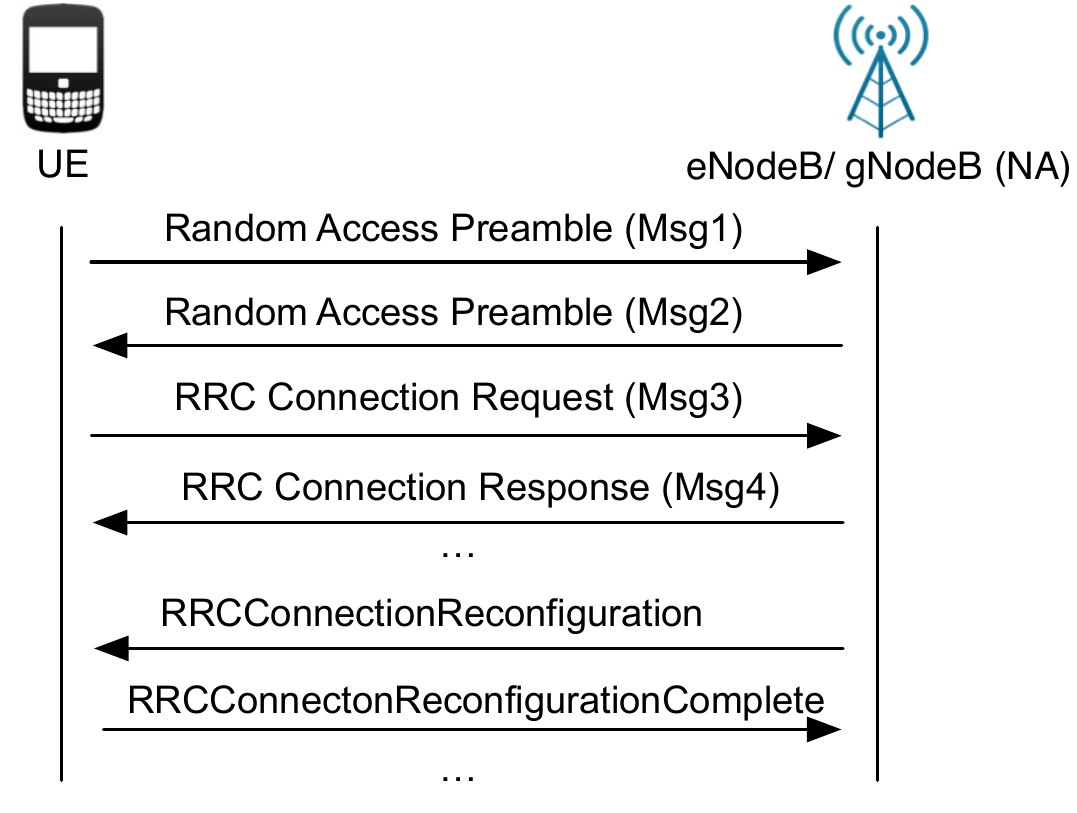}
        \caption{The contention-based random access procedure used in LTE and 5G-SA.}
        \label{fig: lte_5gsa}
    \end{subfigure}
    \hfill
    \begin{subfigure}[t]{0.53\columnwidth}
        \centering
        \includegraphics[height=1in]{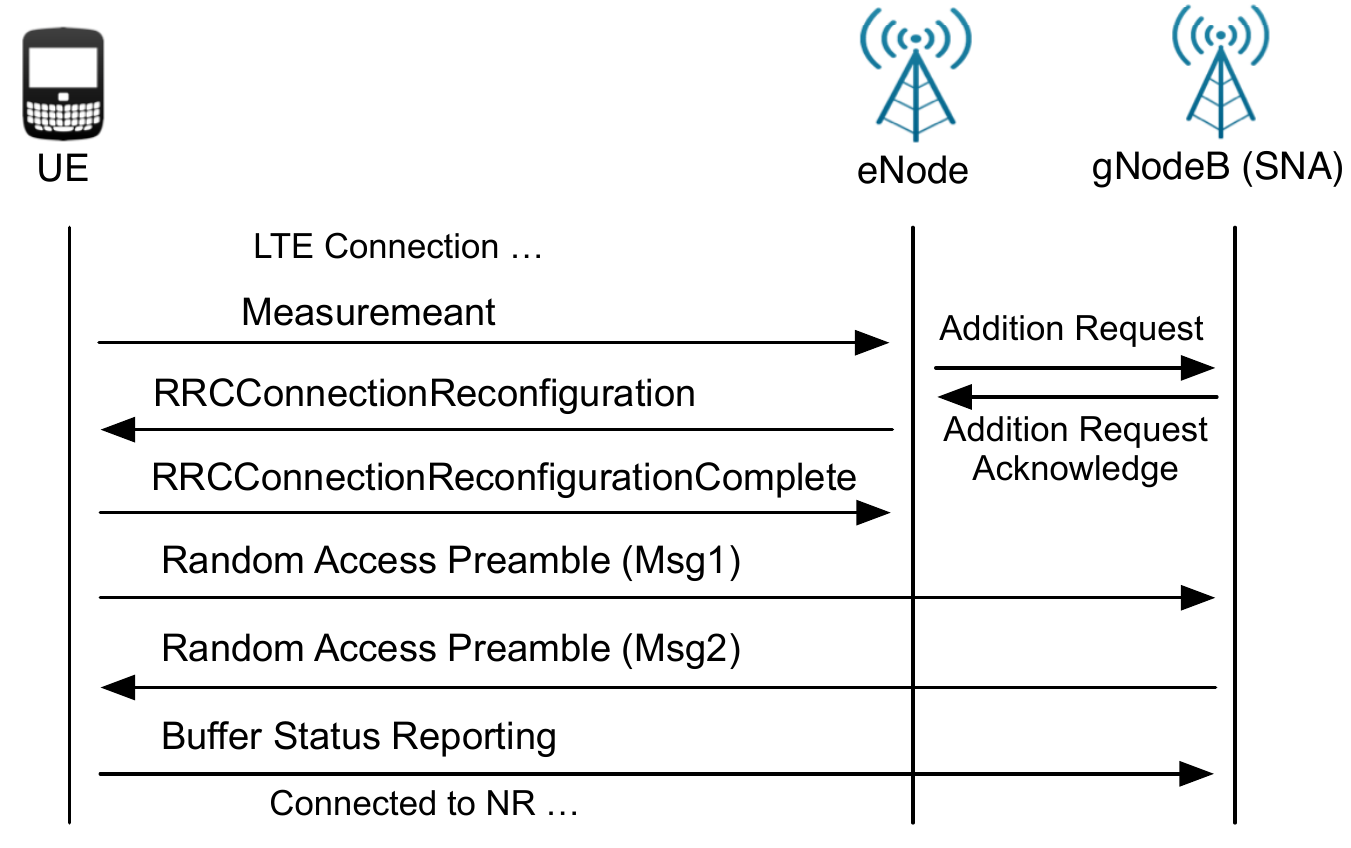}
        \caption{5G radio connection establishment in 5G-NAS.}
        \label{fig: 5gnsa}
    \end{subfigure}
    \caption{Random Access Channel (RACH) procedure as used in LTE/5G-SA(left) and 5G-NSA (right).}
    \label{fig: lte_5g}
\end{figure}

We evaluate the performance of our mobile-relay using a private 5G network deployed with srsRAN~\cite{srsran} and Open5GS~\cite{docker-5gs}. When compared to LTE, 5G provides significant improvements to privacy (e.g., the introduction of concealed identifiers), and bandwidth efficiency (e.g., the addition of native QoS on the SDAP layer). However, these improvements do not prevent the attacks discussed in this paper, with one partial exception which we discuss below.

In 5G, the initial access to the network, i.e. the Random Access Channel (RACH) procedure, can be performed in two ways depending if the network uses a standalone (SA) or a non-standalone (NSA) deployment, Fig.~\ref{fig: lte_5gsa}. The SA version represents the native, efficient 5G procedure. The NSA is a backwards compatible version intended to piggyback on existing 4G/LTE infrastructure. 

\begin{figure}
    \centering
    \begin{subfigure}[b]{\columnwidth}
        \includegraphics[width=0.9\columnwidth]{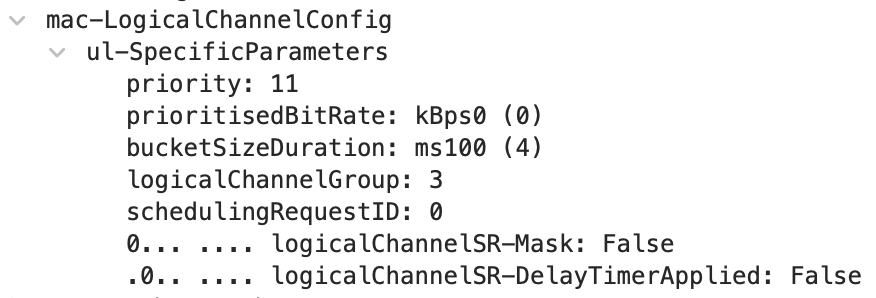}
        \caption{Example of 5G-SA MAC layer configuration inside a \textit{RRCConnectionReconfiguration} message. The \textit{schedulingRequestID} indicates the resource used to send SR messages.}
        \label{fig: 5gsa mac params}
    \end{subfigure}
    \vspace{10pt}
    \begin{subfigure}[b]{\columnwidth}
        \includegraphics[width=0.8\columnwidth]{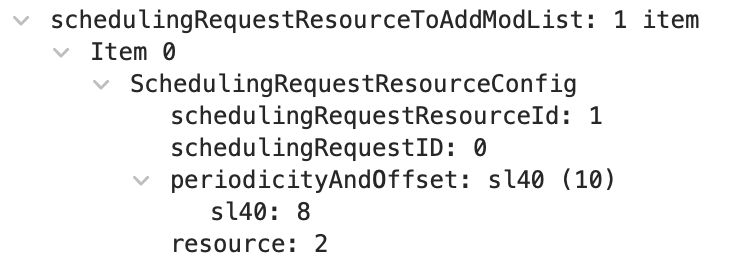}
        \caption{Example of a scheduling request resource, where the \textit{periodictyAndOffset} indicates the periodicity and time slot, and the \textit{resource} indicates the PUCCH index.}
        \label{fig: 5gsa sr group}
    \end{subfigure}
    \caption{\textit{SchedulingRequest} parameters in 5G-SA.}
    \label{fig: 5gsa up params}
\end{figure}

When deploying our relay in a 5G-SA environment we were able to efficiently target the RACH procedure. This is because the initial access to 5G-SA is very similar to LTE in that it uses a contention-based random access channel to initialize the radio connection and configure the default Internet bearer using a \textit{RRCConnectionReconfiguration} message. Thus, our relay is able to begin the guessing procedure when the \textit{RRCConnectionReconfiguration} is observed, wait for scheduling request messages, and compute physical layer parameters using the allocation of NR Physical Uplink Control Channel (NR-PUCCH) values. This process, however, is slightly more difficult in 5G-SA than LTE because LTE follows stricter rules for allocating resource blocks for PUCCH messages~\cite{lin20195g}. We give an example of the 5G-SA SR parameter configuration in  Fig.~\ref{fig: 5gsa up params}. The specific SR resource parameters are configured by \textit{schedulingRequestResourceToAddModlist} which is part of the plain-text \textit{RRCSetup} message. In our 5G-SA experiment, we observe that the gNB does not update these SR parameters when setting up the default Internet bearer. This is expected given that our tests are conducted in a controlled environment, with only one UE connected, which results in conditions that satisfy the latency requirement of Internet bearer and therefore do not require any updates to the SR resource.

Deploying the relay in 5G-NSA setting is significantly more difficult. As shown in Fig.~\ref{fig: 5gnsa}, in 5G-NSA the UE reports signal measurements of surrounding NR cells after being connected to the LTE network. The LTE network can then select a gNodeB station according to the measurements received and request the radio resources on behalf of the UE (e.g., C-RNTI, scheduling request resources) from the gNodeB. Then, the LTE network sends the requested configuration to the UE using a \textit{RRCConnectionReconfiguration} message, and instructs the UE to connect to the gNodeB as a secondary cell. Therefore, the initial access between UE and gNodeB in 5G-NSA uses a contention-free RACH with the preamble parameters indicated in a \textit{RRCConnectionReconfiguration} received from the eNodeB. Additionally, the \textit{RRCConnectionReconfigurationComplete} message is transferred on the established LTE bearer rather than a 5G bearer, which further complicates the problem as no immediate uplink message can be observed by the attacker. As such, maintaining relay radio connections in 5G-NSA is significantly more difficult because: (1) the adversary needs to guess more parameters than in LTE and 5G-SA, such as the preamble parameters and the C-RNTI, and (2) the relay needs to maintain a longer full-spectrum listening window to look for the targeted scheduling request messages. While (1) could be addressed given that the values required are available in other non-encrypted messages, as discussed in Section~\ref{mitigations}, our computationally limited attacker is unable to maintain reliable full spectrum listening windows for sufficient periods in order to address (2).

\section{Discussion}
\subsection{Attack detection} \label{exp imsi catcher detection}

\textit{IMSI-Catcher apps.} We tested the efficiency of IMSI-Catcher apps against our mobile-relay implementation using both a na\"ive self-developed app, which compares the base station reported signal strength with the UE's directly measured one, as well as a 3rd party app i.e. CellularPrivacy~\cite{cellular-privacy}. Our tests were conducted on a Samsung S8 connected through the mobile-relay to Carrier1. Neither app was able to identify our mobile-relay. This is expected, as our passive mobile-relay forwards messages between victim UEs and commercial eNodeBs without any knowledge of cryptographic material. Furthermore, the eNodeB part of the mobile-relay relays valid messages obtained from the commercial eNodeB, making it harder to distinguish between the two. With respect to our self-developed app, we were able to make an interesting observation, namely that the signal strength directly measured by the UE only started to increase significantly for distances less than one meter, which are not realistic from an attacker's perspective.

\noindent \textit{False Base Stations (FBS) detection.} When attempting detection of our active attack (i.e. which is used to obtain the victim's IMSI), we need to modify M-Temporary Mobile Subscriber Identity (M-TMSI) values \textit{once}, which causes either the value itself or the MAC signature of the \textit{Attach Request} message to become invalid and could be, potentially, detectable. However, under normal circumstances, it is common for the \textit{Attach Request} messages to be invalidated in situations such as when the M-TMSI value expires, or when moving to another Mobility Management Entity (MME) group. For this reason, the LTE/5G standard allows multiple re-transmission and corruption of the message itself is not considered malicious.

The 3GPP standard proposes a new potential method for detecting FBSs which uses CRC checksums to verify each physical resource block (Section 6.23~\cite{3gpp-33809}). This allows the network to link specific physical layer messages such as \textit{Scheduling Request} to specific resource blocks. However, this approach is unlikely to fix the underlying causes which enable us to MITM the connection. The relay could easily be modified to ensure that \textit{Uplink Grant} messages, which inform slot allocations, are processed before resource blocks are allocated to the victim UE thus circumventing the benefits of the CRCs.

\subsection{Implications of our work}
In this paper we discuss several attacks that enable an adversary to establish a reliable physical layer MITM position which, in turn, allows them to obtain a victim's identity and recover its VoLTE activity log. Given sufficient hardware resources, an adversary can easily extend our attack to target multiple victims, potentially even located in different geographic areas, simultaneously. We speculate that such an attack could have larger privacy implications, given that such an adversary could correlate call information and determine relationships and activities between these victims simply by using the sequences and timestamps of recovered signalling logs and voice logs.

\subsection{Limitations}
The main limitations of our attack it that it only recovers metadata rather than plaintext such as spoken language or words. While plaintext recovery such as \cite{wright2007language} and \cite{white2011phonotactic} have been shown to work with SIP these do not work with VoLTE/NR. The main reason is that VoLTE/NR uses Adaptive Multi-Rate (AMR) speech coding algorithm instead of the Variable Bit-Rate codec (VBR). The size of VBR coded packet is determined by the encoded audio and thus leaks some information about the encoded payload, however, AMR generates fixed-length packets. Therefore, the choice of using AMR codes in VoLTE/NR represents one of the primary reasons why recognition attacks are limited.

The second significant limitation of our relay is represented by the difficulty to man-in-the-middle LTE Carrier Aggregation (CA) and 5G-NSA connections. Both of these require a relay that supports at least two frequency carriers, a feature that was not available on the B210 SDR. Another related issue is the contention-free RACH procedure which uses \textit{RRCConnectionReconfiguration} encrypted messages to relay physical layer parameters to the UE and which increases the difficulty of obtaining these 5G-NSA networks.

\subsection{Attack mitigations and defences}
\label{mitigations}
Attack mitigations and defences for the proposed work fall in two main categories: (1) preventing VoLTE traffic identification and (2) increasing the difficulty of deploying the mobile-relay. 

As stated previously, VoLTE sequence recovery mainly relies on using metadata such as message length and type to identify messages. Plaintext padding techniques could help mitigate the problem to some extent, however they would not be advisable in a mobile communication scenario due to the significant impact on bandwidth. For example, when using the Samsung S7 UE with Carrier1, the maximum, average, and minimum uplink VoLTE message lengths are 2479, 1170, and 337 bytes, respectively (see Table~\ref{tab:signal_len_map}). In order to achieve the best protection, padding for all messages would need to be done to the maximum size (e.g., 2479B) however this would result in an uplink bandwidth drop of about $48.5\%$. Disabling Voice Activity Detection (VAD) prevents the attacker from learning voice activity information, however, it results in significant waste of bandwidth and spectrum resources. For example, with VAD enabled a one-minute VoLTE call between Alice and Bob with $50\%$ voice saturation generates 1687 uplink RTP packets. With VAD disabled the same call generates 3000 uplink packets representing a $77.8\%$ increase.

The key method for preventing the mobile-relay deployment is to increase the difficulty of guessing physical layer parameters. First, we can randomize the \textit{sr-PUCCH-ResourceIndex} and decrease the value of \textit{dsr-TranMax}. However, the LTE PUCCH is located at the edge of carrier bandwidth~\cite{3gpp-36211} (Section 5.4.3), therefore, the option for \textit{sr-PUCCH-ResourceIndex} is limited. As introduced in Section~\ref{volte-related-packets}, we need at least one scheduling request message to calculate physical layer parameters, therefore setting \textit{dsr-TranMax} to 1 can hinder this computation. Lower values for \textit{dsr-TranMax} do have implications for the robustness of the network in poor signal circumstances (e.g., when the UE is behind walls, or is far away from the base station). Another possibility is to increase the time window between receiving \textit{RRCConnectionReconfiguration} and sending \textit{RRCConnectionReconfigurationCompelete} messages, which complicates guessing by extending the search window. However, this window extension increases the possibility of radio signal interference (see Section~\ref{exp: guessing physical layer parameters}).

As such, we believe that a slightly modified version of 5G-NSA, described in the following, is most likely to be efficient against our physical layer relay. First, a successfully deployed relay needs to obtain the physical layer parameters from the \textit{Scheduling Request (SR)} messages. Then, the attacker also requires knowledge about the victim's C-RNTI identity in order to select the correct downlink messages to be forwarded to the target UE. As discussed in Section~\ref{5G-eval}, in the 5G-NSA attachment procedure these specific parameters are sent to the UE inside an encrypted \textit{RRCConnectionReconfiguration} message which makes the attack more difficult, it requires an extended listening window for capturing the \textit{SR} message, and forces the attacker to recover the new, 5G C-RNTI value from a different message, i.e. the \textit{BufferStatusReporting (BSR)}. While protecting the \textit{SR} is not possible as it contains low level configuration for the physical layer which needs to be directly available to the UE, the C-RNTI could be. One relatively straight-forward method would involve two minor alterations to the 5G-NSA procedure. First, a new security context should be established on the 5G C-RNTI, instead of only temporarily relying on it to facilitate the contention-free RACH. Second the 5G C-RNTI needs to be kept secret, thus it should not be transmitted inside MAC layer messages such as \textit{BSR}, but instead should be moved on to the RRC layer. We believe that these changes would significantly reduce the attack surface, however, they represent significant changes to procedures in both 5G and LTE standards and therefore would require extensive testing on specialized prototype infrastructure which goes beyond the purpose of this work.

\subsection{Ethical Considerations}
In developing and evaluating our attacks, we comply with the law and other users' privacy by controlling the transmission powers of our mobile-relay in order to avoid attracting neighbouring UEs and cause interference with commercial eNodeBs.

\section{Conclusion}
While a lot of privacy related research in LTE and 5G is focused on the radio interface, VoLTE/NR privacy has remained largely unexplored. In this work, we showed two types of privacy attacks: a VoLTE/NR activity monitoring attack, which exploits encrypted PDCP data and recovers VoLTE/NR activities, and an identity recovery attack, which is able to obtain and link network identifiers to victims' phone numbers using VoLTE/NR traffic. We also proposed and implemented several improvements to the relay attacker, which greatly improve its undetectability and reliability. We have further shown the real-world performance of our attacks by recovering victims' VoLTE/NR activity logs from the encrypted traffic collected, and then linking their anonymised identifiers to their real-life correspondents. Finally, we conclude by providing a discussion on the mitigations and defense for the proposed attacks.

\begin{acks}
This work is partially funded by the China Scholarship Council (CSC) with awards to Zishuai Cheng, and Engineering and Physical Sciences Research Council (EPSRC) under grants EP/R012598/1, 	EP/R008000/1 and EP/V000454/1.
\end{acks}

\bibliographystyle{ACM-Reference-Format}
\bibliography{refs}


\begin{thebibliography}{37}


\ifx \showCODEN    \undefined \def \showCODEN     #1{\unskip}     \fi
\ifx \showDOI      \undefined \def \showDOI       #1{#1}\fi
\ifx \showISBNx    \undefined \def \showISBNx     #1{\unskip}     \fi
\ifx \showISBNxiii \undefined \def \showISBNxiii  #1{\unskip}     \fi
\ifx \showISSN     \undefined \def \showISSN      #1{\unskip}     \fi
\ifx \showLCCN     \undefined \def \showLCCN      #1{\unskip}     \fi
\ifx \shownote     \undefined \def \shownote      #1{#1}          \fi
\ifx \showarticletitle \undefined \def \showarticletitle #1{#1}   \fi
\ifx \showURL      \undefined \def \showURL       {\relax}        \fi
\providecommand\bibfield[2]{#2}
\providecommand\bibinfo[2]{#2}
\providecommand\natexlab[1]{#1}
\providecommand\showeprint[2][]{arXiv:#2}

\bibitem[3GPP(2022a)]%
        {3gpp-23203}
\bibfield{author}{\bibinfo{person}{3GPP}.} \bibinfo{year}{2022}\natexlab{a}.
\newblock \bibinfo{title}{3GPP 23.203: Policy and charging control
  architecture}.
\newblock
\newblock
\urldef\tempurl%
\url{https://portal.3gpp.org/desktopmodules/Specifications/SpecificationDetails.aspx?specificationId=810}
\showURL{%
\tempurl}
\newblock
\shownote{[Online; accessed 20-May-2022]}.


\bibitem[3GPP(2022b)]%
        {3gpp-24301}
\bibfield{author}{\bibinfo{person}{3GPP}.} \bibinfo{year}{2022}\natexlab{b}.
\newblock \bibinfo{title}{3GPP 24.301: Non-Access-Stratum (NAS) protocol for
  Evolved Packet System (EPS); Stage 3}.
\newblock
\newblock
\urldef\tempurl%
\url{https://portal.3gpp.org/desktopmodules/Specifications/SpecificationDetails.aspx?specificationId=1072}
\showURL{%
\tempurl}
\newblock
\shownote{[Online; accessed 30-May-2022]}.


\bibitem[3GPP(2022c)]%
        {3gpp-26071}
\bibfield{author}{\bibinfo{person}{3GPP}.} \bibinfo{year}{2022}\natexlab{c}.
\newblock \bibinfo{title}{3GPP 26.071: Mandatory speech CODEC speech processing
  functions; AMR speech Codec; General description}.
\newblock
\newblock
\urldef\tempurl%
\url{https://portal.3gpp.org/desktopmodules/Specifications/SpecificationDetails.aspx?specificationId=1386}
\showURL{%
\tempurl}
\newblock
\shownote{[Online; accessed 16-Mar-2022]}.


\bibitem[3GPP(2022d)]%
        {3gpp-26090}
\bibfield{author}{\bibinfo{person}{3GPP}.} \bibinfo{year}{2022}\natexlab{d}.
\newblock \bibinfo{title}{3GPP 26.090: Mandatory Speech Codec speech processing
  functions; Adaptive Multi-Rate (AMR) speech codec; Transcoding functions}.
\newblock
\newblock
\urldef\tempurl%
\url{https://portal.3gpp.org/desktopmodules/Specifications/SpecificationDetails.aspx?specificationId=1392}
\showURL{%
\tempurl}
\newblock
\shownote{[Online; accessed 18-Mar-2022]}.


\bibitem[3GPP(2022e)]%
        {3gpp-26201}
\bibfield{author}{\bibinfo{person}{3GPP}.} \bibinfo{year}{2022}\natexlab{e}.
\newblock \bibinfo{title}{3GPP 26.201: Speech codec speech processing
  functions; Adaptive Multi-Rate - Wideband (AMR-WB) speech codec; Frame
  structure}.
\newblock
\newblock
\urldef\tempurl%
\url{https://portal.3gpp.org/desktopmodules/Specifications/SpecificationDetails.aspx?specificationId=1429}
\showURL{%
\tempurl}
\newblock
\shownote{[Online; accessed 18-Mar-2022]}.


\bibitem[3GPP(2022f)]%
        {3gpp-36213}
\bibfield{author}{\bibinfo{person}{3GPP}.} \bibinfo{year}{2022}\natexlab{f}.
\newblock \bibinfo{title}{3GPP 36.213: Evolved Universal Terrestrial Radio
  Access (E-UTRA); Physical layer procedures}.
\newblock
\newblock
\urldef\tempurl%
\url{https://portal.3gpp.org/desktopmodules/Specifications/SpecificationDetails.aspx?specificationId=2427}
\showURL{%
\tempurl}
\newblock
\shownote{[Online; accessed 30-May-2022]}.


\bibitem[3GPP(2022g)]%
        {3gpp-36321}
\bibfield{author}{\bibinfo{person}{3GPP}.} \bibinfo{year}{2022}\natexlab{g}.
\newblock \bibinfo{title}{Evolved Universal Terrestrial Radio Access (E-UTRA);
  Medium Access Control (MAC) protocol specification}.
\newblock
\newblock
\urldef\tempurl%
\url{https://portal.3gpp.org/desktopmodules/Specifications/SpecificationDetails.aspx?specificationId=2437}
\showURL{%
\tempurl}
\newblock
\shownote{[Online; accessed 30-May-2022]}.


\bibitem[3GPP(2022h)]%
        {3gpp-36323}
\bibfield{author}{\bibinfo{person}{3GPP}.} \bibinfo{year}{2022}\natexlab{h}.
\newblock \bibinfo{title}{Evolved Universal Terrestrial Radio Access (E-UTRA);
  Packet Data Convergence Protocol (PDCP) specification}.
\newblock
\newblock
\urldef\tempurl%
\url{https://portal.3gpp.org/desktopmodules/Specifications/SpecificationDetails.aspx?specificationId=2439}
\showURL{%
\tempurl}
\newblock
\shownote{[Online; accessed 30-May-2022]}.


\bibitem[3GPP(2022i)]%
        {3gpp-36211}
\bibfield{author}{\bibinfo{person}{3GPP}.} \bibinfo{year}{2022}\natexlab{i}.
\newblock \bibinfo{title}{Evolved Universal Terrestrial Radio Access (E-UTRA);
  Physical channels and modulation}.
\newblock
\newblock
\urldef\tempurl%
\url{https://portal.3gpp.org/desktopmodules/Specifications/SpecificationDetails.aspx?specificationId=2425}
\showURL{%
\tempurl}
\newblock
\shownote{[Online; accessed 30-May-2022]}.


\bibitem[3GPP(2022j)]%
        {3gpp-26102}
\bibfield{author}{\bibinfo{person}{3GPP}.} \bibinfo{year}{2022}\natexlab{j}.
\newblock \bibinfo{title}{Mandatory speech codec; Adaptive Multi-Rate (AMR)
  speech codec; Interface to Iu, Uu and Nb}.
\newblock
\newblock
\urldef\tempurl%
\url{https://portal.3gpp.org/desktopmodules/Specifications/SpecificationDetails.aspx?specificationId=1398}
\showURL{%
\tempurl}
\newblock
\shownote{[Online; accessed 30-May-2022]}.


\bibitem[3GPP(2022k)]%
        {3gpp-33809}
\bibfield{author}{\bibinfo{person}{3GPP}.} \bibinfo{year}{2022}\natexlab{k}.
\newblock \bibinfo{title}{Study on 5G security enhancements against False Base
  Stations (FBS)}.
\newblock
\newblock
\urldef\tempurl%
\url{https://portal.3gpp.org/desktopmodules/Specifications/SpecificationDetails.aspx?specificationId=3539}
\showURL{%
\tempurl}
\newblock
\shownote{[Online; accessed 10-Aug-2022]}.


\bibitem[Apple(2022)]%
        {rvi}
\bibfield{author}{\bibinfo{person}{Apple}.} \bibinfo{year}{2022}\natexlab{}.
\newblock \bibinfo{title}{Recording a Packet Trace}.
\newblock
\newblock
\urldef\tempurl%
\url{https://developer.apple.com/documentation/network/recording\_a\_packet\_trace}
\showURL{%
\tempurl}
\newblock
\shownote{[Online; accessed 20-May-2022]}.


\bibitem[Bae et~al\mbox{.}(2022)]%
        {277230}
\bibfield{author}{\bibinfo{person}{Sangwook Bae}, \bibinfo{person}{Mincheol
  Son}, \bibinfo{person}{Dongkwan Kim}, \bibinfo{person}{CheolJun Park},
  \bibinfo{person}{Jiho Lee}, \bibinfo{person}{Sooel Son}, {and}
  \bibinfo{person}{Yongdae Kim}.} \bibinfo{year}{2022}\natexlab{}.
\newblock \showarticletitle{Watching the Watchers: Practical Video
  Identification Attack in {LTE} Networks}. In \bibinfo{booktitle}{\emph{31st
  USENIX Security Symposium (USENIX Security 22)}}. \bibinfo{publisher}{USENIX
  Association}, \bibinfo{address}{Boston, MA}, \bibinfo{pages}{1307--1324}.
\newblock
\showISBNx{978-1-939133-31-1}
\urldef\tempurl%
\url{https://www.usenix.org/conference/usenixsecurity22/presentation/bae}
\showURL{%
\tempurl}


\bibitem[CellularPrivacy(2022)]%
        {cellular-privacy}
\bibfield{author}{\bibinfo{person}{CellularPrivacy}.}
  \bibinfo{year}{2022}\natexlab{}.
\newblock \bibinfo{title}{Android-IMSI-Catcher-Detector}.
\newblock
\newblock
\urldef\tempurl%
\url{https://github.com/CellularPrivacy/Android-IMSI-Catcher-Detector}
\showURL{%
\tempurl}
\newblock
\shownote{[Online; accessed 10-Aug-2022]}.


\bibitem[Chlosta et~al\mbox{.}(2021)]%
        {chlosta20215g}
\bibfield{author}{\bibinfo{person}{Merlin Chlosta}, \bibinfo{person}{David
  Rupprecht}, \bibinfo{person}{Christina P\"{o}pper}, {and}
  \bibinfo{person}{Thorsten Holz}.} \bibinfo{year}{2021}\natexlab{}.
\newblock \showarticletitle{5G SUCI-Catchers: Still Catching Them All?}. In
  \bibinfo{booktitle}{\emph{Proceedings of the 14th ACM Conference on Security
  and Privacy in Wireless and Mobile Networks}} (Abu Dhabi, United Arab
  Emirates) \emph{(\bibinfo{series}{WiSec '21})}.
  \bibinfo{publisher}{Association for Computing Machinery},
  \bibinfo{address}{New York, NY, USA}, \bibinfo{pages}{359–364}.
\newblock
\showISBNx{9781450383493}
\urldef\tempurl%
\url{https://doi.org/10.1145/3448300.3467826}
\showDOI{\tempurl}


\bibitem[Erni et~al\mbox{.}(2022)]%
        {arxiv.2106.05039}
\bibfield{author}{\bibinfo{person}{Simon Erni}, \bibinfo{person}{Martin
  Kotuliak}, \bibinfo{person}{Patrick Leu}, \bibinfo{person}{Marc Roeschlin},
  {and} \bibinfo{person}{Srdjan Capkun}.} \bibinfo{year}{2022}\natexlab{}.
\newblock \showarticletitle{AdaptOver: Adaptive Overshadowing Attacks in
  Cellular Networks}. In \bibinfo{booktitle}{\emph{Proceedings of the 28th
  Annual International Conference on Mobile Computing And Networking}} (Sydney,
  NSW, Australia) \emph{(\bibinfo{series}{MobiCom '22})}.
  \bibinfo{publisher}{Association for Computing Machinery},
  \bibinfo{address}{New York, NY, USA}, \bibinfo{pages}{743–--755}.
\newblock
\showISBNx{9781450391818}
\urldef\tempurl%
\url{https://doi.org/10.1145/3495243.3560525}
\showDOI{\tempurl}


\bibitem[GSMA(2022)]%
        {the-mobile-economy}
\bibfield{author}{\bibinfo{person}{GSMA}.} \bibinfo{year}{2022}\natexlab{}.
\newblock \bibinfo{title}{The Mobile Economy}.
\newblock
\newblock
\urldef\tempurl%
\url{https://www.gsma.com/mobileeconomy/wp-content/uploads/2022/02/280222-The-Mobile-Economy-2022.pdf}
\showURL{%
\tempurl}
\newblock
\shownote{[Online; accessed 30-May-2022]}.


\bibitem[Herle(2022)]%
        {docker-5gs}
\bibfield{author}{\bibinfo{person}{Supreeth Herle}.}
  \bibinfo{year}{2022}\natexlab{}.
\newblock \bibinfo{title}{Docker Open5GS}.
\newblock
\newblock
\urldef\tempurl%
\url{https://github.com/herlesupreeth/docker\_open5gs}
\showURL{%
\tempurl}
\newblock
\shownote{[Online; accessed 30-May-2022]}.


\bibitem[Hong et~al\mbox{.}(2018a)]%
        {hong2018guti}
\bibfield{author}{\bibinfo{person}{Byeongdo Hong}, \bibinfo{person}{Sangwook
  Bae}, {and} \bibinfo{person}{Yongdae Kim}.} \bibinfo{year}{2018}\natexlab{a}.
\newblock \showarticletitle{GUTI Reallocation Demystified: Cellular Location
  Tracking with Changing Temporary Identifier.}. In
  \bibinfo{booktitle}{\emph{NDSS}} (San Diego, CA, USA).
\newblock


\bibitem[Hong et~al\mbox{.}(2018b)]%
        {hong2018peeking}
\bibfield{author}{\bibinfo{person}{Byeongdo Hong}, \bibinfo{person}{Shinjo
  Park}, \bibinfo{person}{Hongil Kim}, \bibinfo{person}{Dongkwan Kim},
  \bibinfo{person}{Hyunwook Hong}, \bibinfo{person}{Hyunwoo Choi},
  \bibinfo{person}{Jean-Pierre Seifert}, \bibinfo{person}{Sung-Ju Lee}, {and}
  \bibinfo{person}{Yongdae Kim}.} \bibinfo{year}{2018}\natexlab{b}.
\newblock \showarticletitle{Peeking over the cellular walled gardens-a method
  for closed network diagnosis}.
\newblock \bibinfo{journal}{\emph{IEEE Transactions on Mobile Computing}}
  \bibinfo{volume}{17}, \bibinfo{number}{10} (\bibinfo{year}{2018}),
  \bibinfo{pages}{2366--2380}.
\newblock


\bibitem[Kim et~al\mbox{.}(2015)]%
        {kim2015breaking}
\bibfield{author}{\bibinfo{person}{Hongil Kim}, \bibinfo{person}{Dongkwan Kim},
  \bibinfo{person}{Minhee Kwon}, \bibinfo{person}{Hyungseok Han},
  \bibinfo{person}{Yeongjin Jang}, \bibinfo{person}{Dongsu Han},
  \bibinfo{person}{Taesoo Kim}, {and} \bibinfo{person}{Yongdae Kim}.}
  \bibinfo{year}{2015}\natexlab{}.
\newblock \showarticletitle{Breaking and Fixing VoLTE: Exploiting Hidden Data
  Channels and Mis-Implementations}. In \bibinfo{booktitle}{\emph{Proceedings
  of the 22nd ACM SIGSAC Conference on Computer and Communications Security}}
  (Denver, Colorado, USA) \emph{(\bibinfo{series}{CCS '15})}.
  \bibinfo{publisher}{Association for Computing Machinery},
  \bibinfo{address}{New York, NY, USA}, \bibinfo{pages}{328–339}.
\newblock
\showISBNx{9781450338325}
\urldef\tempurl%
\url{https://doi.org/10.1145/2810103.2813718}
\showDOI{\tempurl}


\bibitem[Kohls et~al\mbox{.}(2019)]%
        {kohls-lost-2019}
\bibfield{author}{\bibinfo{person}{Katharina Kohls}, \bibinfo{person}{David
  Rupprecht}, \bibinfo{person}{Thorsten Holz}, {and} \bibinfo{person}{Christina
  P\"{o}pper}.} \bibinfo{year}{2019}\natexlab{}.
\newblock \showarticletitle{Lost Traffic Encryption: Fingerprinting LTE/4G
  Traffic on Layer Two}. In \bibinfo{booktitle}{\emph{Proceedings of the 12th
  Conference on Security and Privacy in Wireless and Mobile Networks}} (Miami,
  Florida) \emph{(\bibinfo{series}{WiSec '19})}.
  \bibinfo{publisher}{Association for Computing Machinery},
  \bibinfo{address}{New York, NY, USA}, \bibinfo{pages}{249–260}.
\newblock
\showISBNx{9781450367264}
\urldef\tempurl%
\url{https://doi.org/10.1145/3317549.3323416}
\showDOI{\tempurl}


\bibitem[Kotuliak et~al\mbox{.}(2022)]%
        {kotuliak2022ltrack}
\bibfield{author}{\bibinfo{person}{Martin Kotuliak}, \bibinfo{person}{Simon
  Erni}, \bibinfo{person}{Patrick Leu}, \bibinfo{person}{Marc R{\"o}schlin},
  {and} \bibinfo{person}{Srdjan Capkun}.} \bibinfo{year}{2022}\natexlab{}.
\newblock \showarticletitle{{LTrack}: Stealthy Tracking of Mobile Phones in
  {LTE}}. In \bibinfo{booktitle}{\emph{31st USENIX Security Symposium (USENIX
  Security 22)}}. \bibinfo{publisher}{USENIX Association},
  \bibinfo{address}{Boston, MA}, \bibinfo{pages}{1291--1306}.
\newblock
\showISBNx{978-1-939133-31-1}
\urldef\tempurl%
\url{https://www.usenix.org/conference/usenixsecurity22/presentation/kotuliak}
\showURL{%
\tempurl}


\bibitem[Kune et~al\mbox{.}(2012)]%
        {kune2012location}
\bibfield{author}{\bibinfo{person}{Denis~Foo Kune}, \bibinfo{person}{John
  Koelndorfer}, \bibinfo{person}{Nicholas Hopper}, {and}
  \bibinfo{person}{Yongdae Kim}.} \bibinfo{year}{2012}\natexlab{}.
\newblock \showarticletitle{Location leaks on the GSM air interface}.
\newblock \bibinfo{journal}{\emph{Network and Distributed Systems Security
  (NDSS) Symposium2012}} (\bibinfo{year}{2012}).
\newblock


\bibitem[Lin et~al\mbox{.}(2019)]%
        {lin20195g}
\bibfield{author}{\bibinfo{person}{Xingqin Lin}, \bibinfo{person}{Jingya Li},
  \bibinfo{person}{Robert Baldemair}, \bibinfo{person}{Jung-Fu~Thomas Cheng},
  \bibinfo{person}{Stefan Parkvall}, \bibinfo{person}{Daniel~Chen Larsson},
  \bibinfo{person}{Havish Koorapaty}, \bibinfo{person}{Mattias Frenne},
  \bibinfo{person}{Sorour Falahati}, \bibinfo{person}{Asbjorn Grovlen},
  {et~al\mbox{.}}} \bibinfo{year}{2019}\natexlab{}.
\newblock \showarticletitle{5G new radio: Unveiling the essentials of the next
  generation wireless access technology}.
\newblock \bibinfo{journal}{\emph{IEEE Communications Standards Magazine}}
  \bibinfo{volume}{3}, \bibinfo{number}{3} (\bibinfo{year}{2019}),
  \bibinfo{pages}{30--37}.
\newblock


\bibitem[Lu et~al\mbox{.}(2020)]%
        {lu2020ghost}
\bibfield{author}{\bibinfo{person}{Yu-Han Lu}, \bibinfo{person}{Chi-Yu Li},
  \bibinfo{person}{Yao-Yu Li}, \bibinfo{person}{Sandy Hsin-Yu Hsiao},
  \bibinfo{person}{Tian Xie}, \bibinfo{person}{Guan-Hua Tu}, {and}
  \bibinfo{person}{Wei-Xun Chen}.} \bibinfo{year}{2020}\natexlab{}.
\newblock \showarticletitle{Ghost Calls from Operational 4G Call Systems: IMS
  Vulnerability, Call DoS Attack, and Countermeasure}. In
  \bibinfo{booktitle}{\emph{Proceedings of the 26th Annual International
  Conference on Mobile Computing and Networking}} (London, United Kingdom)
  \emph{(\bibinfo{series}{MobiCom '20})}. \bibinfo{publisher}{Association for
  Computing Machinery}, \bibinfo{address}{New York, NY, USA}, Article
  \bibinfo{articleno}{8}, \bibinfo{numpages}{14}~pages.
\newblock
\showISBNx{9781450370851}
\urldef\tempurl%
\url{https://doi.org/10.1145/3372224.3380885}
\showDOI{\tempurl}


\bibitem[Osmocom(2022)]%
        {simtrace}
\bibfield{author}{\bibinfo{person}{Osmocom}.} \bibinfo{year}{2022}\natexlab{}.
\newblock \bibinfo{title}{SIMtrace 2}.
\newblock
\newblock
\urldef\tempurl%
\url{https://osmocom.org/projects/simtrace2/wiki}
\showURL{%
\tempurl}
\newblock
\shownote{[Online; accessed 20-May-2022]}.


\bibitem[RFC(2022)]%
        {rfc-3095}
\bibfield{author}{\bibinfo{person}{RFC}.} \bibinfo{year}{2022}\natexlab{}.
\newblock \bibinfo{title}{RObust Header Compression (ROHC): Framework and four
  profiles: RTP, UDP, ESP, and uncompressed}.
\newblock
\newblock
\urldef\tempurl%
\url{https://datatracker.ietf.org/doc/html/rfc3095}
\showURL{%
\tempurl}
\newblock
\shownote{[Online; accessed 20-May-2022]}.


\bibitem[Rupprecht et~al\mbox{.}(2019)]%
        {rupprecht-breaking-2019}
\bibfield{author}{\bibinfo{person}{David Rupprecht}, \bibinfo{person}{Katharina
  Kohls}, \bibinfo{person}{Thorsten Holz}, {and} \bibinfo{person}{Christina
  Popper}.} \bibinfo{year}{2019}\natexlab{}.
\newblock \showarticletitle{Breaking {LTE} on Layer Two}. In
  \bibinfo{booktitle}{\emph{2019 {IEEE} Symposium on Security and Privacy
  ({SP})}} (San Francisco, {CA}, {USA}, 2019-05). \bibinfo{publisher}{{IEEE}},
  \bibinfo{pages}{1121--1136}.
\newblock
\showISBNx{978-1-5386-6660-9}


\bibitem[Rupprecht et~al\mbox{.}(2020a)]%
        {rupprecht2020imp4gt}
\bibfield{author}{\bibinfo{person}{David Rupprecht}, \bibinfo{person}{Katharina
  Kohls}, \bibinfo{person}{Thorsten Holz}, {and} \bibinfo{person}{Christina
  P{\"o}pper}.} \bibinfo{year}{2020}\natexlab{a}.
\newblock \showarticletitle{IMP4GT: IMPersonation Attacks in 4G NeTworks.}. In
  \bibinfo{booktitle}{\emph{ISOC Network and Distributed System Security
  Symposium (NDSS)}} (San Diego, CA, USA). \bibinfo{publisher}{ISOC}.
\newblock


\bibitem[Rupprecht et~al\mbox{.}(2020b)]%
        {rupprecht-call-2020}
\bibfield{author}{\bibinfo{person}{David Rupprecht}, \bibinfo{person}{Katharina
  Kohls}, \bibinfo{person}{Christina Pöpper}, {and} \bibinfo{person}{Thorsten
  Holz}.} \bibinfo{year}{2020}\natexlab{b}.
\newblock \showarticletitle{Call Me Maybe: Eavesdropping Encrypted {LTE} Calls
  With {ReVoLTE}}. In \bibinfo{booktitle}{\emph{29th {USENIX} Security
  Symposium ({USENIX} Security 20)}} (2020). \bibinfo{publisher}{{USENIX}
  Association}, \bibinfo{pages}{73--88}.
\newblock
\showISBNx{978-1-939133-17-5}


\bibitem[Shaik et~al\mbox{.}(2019)]%
        {shaik-practical-2015}
\bibfield{author}{\bibinfo{person}{Altaf Shaik}, \bibinfo{person}{Ravishankar
  Borgaonkar}, \bibinfo{person}{N Asokan}, \bibinfo{person}{Valtteri Niemi},
  {and} \bibinfo{person}{Jean-Pierre Seifert}.}
  \bibinfo{year}{2019}\natexlab{}.
\newblock \showarticletitle{Practical attacks against privacy and availability
  in 4G/LTE mobile communication systems}.
\newblock


\bibitem[Systems(2022)]%
        {srsran}
\bibfield{author}{\bibinfo{person}{Software~Radio Systems}.}
  \bibinfo{year}{2022}\natexlab{}.
\newblock \bibinfo{title}{Open source SDR 4G/5G software suite from Software
  Radio Systems (SRS)}.
\newblock
\newblock
\urldef\tempurl%
\url{https://github.com/srsran/srsRAN}
\showURL{%
\tempurl}
\newblock
\shownote{[Online; accessed 20-May-2022]}.


\bibitem[White et~al\mbox{.}(2011)]%
        {white2011phonotactic}
\bibfield{author}{\bibinfo{person}{Andrew~M White}, \bibinfo{person}{Austin~R
  Matthews}, \bibinfo{person}{Kevin~Z Snow}, {and} \bibinfo{person}{Fabian
  Monrose}.} \bibinfo{year}{2011}\natexlab{}.
\newblock \showarticletitle{Phonotactic reconstruction of encrypted voip
  conversations: Hookt on fon-iks}. In \bibinfo{booktitle}{\emph{2011 IEEE
  Symposium on Security and Privacy}} (USA). IEEE, \bibinfo{publisher}{IEEE},
  \bibinfo{pages}{3--18}.
\newblock


\bibitem[Wright et~al\mbox{.}(2007)]%
        {wright2007language}
\bibfield{author}{\bibinfo{person}{Charles~V Wright}, \bibinfo{person}{Lucas
  Ballard}, \bibinfo{person}{Fabian Monrose}, {and} \bibinfo{person}{Gerald~M
  Masson}.} \bibinfo{year}{2007}\natexlab{}.
\newblock \showarticletitle{Language identification of encrypted voip traffic:
  Alejandra y roberto or alice and bob?}. In \bibinfo{booktitle}{\emph{USENIX
  Security Symposium}}, Vol.~\bibinfo{volume}{3}. \bibinfo{publisher}{USENIX
  Association}, \bibinfo{address}{Boston, MA}, \bibinfo{pages}{43--54}.
\newblock


\bibitem[Xie et~al\mbox{.}(2018)]%
        {8433136}
\bibfield{author}{\bibinfo{person}{Tian Xie}, \bibinfo{person}{Guan-Hua Tu},
  \bibinfo{person}{Chi-Yu Li}, \bibinfo{person}{Chunyi Peng},
  \bibinfo{person}{Jiawei Li}, {and} \bibinfo{person}{Mi Zhang}.}
  \bibinfo{year}{2018}\natexlab{}.
\newblock \showarticletitle{The Dark Side of Operational Wi-Fi Calling
  Services}. In \bibinfo{booktitle}{\emph{2018 IEEE Conference on
  Communications and Network Security (CNS)}} (Beijing, China).
  \bibinfo{publisher}{IEEE}, \bibinfo{pages}{1--1}.
\newblock
\urldef\tempurl%
\url{https://doi.org/10.1109/CNS.2018.8433136}
\showDOI{\tempurl}


\bibitem[Yang et~al\mbox{.}(2019)]%
        {yang2019hiding}
\bibfield{author}{\bibinfo{person}{Hojoon Yang}, \bibinfo{person}{Sangwook
  Bae}, \bibinfo{person}{Mincheol Son}, \bibinfo{person}{Hongil Kim},
  \bibinfo{person}{Song~Min Kim}, {and} \bibinfo{person}{Yongdae Kim}.}
  \bibinfo{year}{2019}\natexlab{}.
\newblock \showarticletitle{Hiding in plain signal: Physical signal
  overshadowing attack on $\{$LTE$\}$}. In \bibinfo{booktitle}{\emph{28th
  USENIX Security Symposium (USENIX Security 19)}}. \bibinfo{publisher}{USENIX
  Association}, \bibinfo{address}{Boston, MA}, \bibinfo{pages}{55--72}.
\newblock


\end{thebibliography}

\appendix

\section*{APPENDIX}

\section{Algorithms}\label{appendix:al}

\begin{algorithm}[h!]
    \scriptsize
    \caption{\textit{schedulingRequestConfig} computation}
    \label{guessing_scheduling_request_parameters}
    \SetKwFunction{AnlysSRParams}{AnalyseSRParameters}
    
    \KwInput{\textit{rnti},\textit{p}}
    \KwOutput{\textit{sr-ConfigIndex},\textit{sr-PUCCH-ResourceIndex}}
    \BlankLine
    
    \Fn{\AnlysSRParams{\textit{rnti}, \textit{p}}}{
        \textbf{mobile-relay}: open all slots $S$ and sub-carriers $C$ for \textit{rnti}\\
        \For {$sr \in S\times C$ \textbf{and} \textit{rnti}}{
            \uIf{$sr$ \textbf{is} first request \textbf{and} \textit{p} $= 0$}{
                ${tti}'_{sr} \xleftarrow{}$ \textit{10 $\cdot$ system frame number + subframe number}\\ 
                \textbf{flush} $sr$\\
            }\uElseIf{\textit{p} $\neq 0$}{
                \Goto \ref{CalSRParmas}
            }
            \Else{
                ${tti}''_{sr} \xleftarrow{}$ \textit{10 $\cdot$ system frame number + subframe number}\\ 
                \eIf{${tti}''_{sr} > {tti}'_{sr}$}{
                    $p \xleftarrow{} {tti}''_{sr} - {tti}'_{sr}$
                }{
                    $p \xleftarrow{} {tti}''_{sr} + 1024 - {tti}'_{sr}$
                }
            }
            
            \textit{subfrm-off} $\xleftarrow{} {tti}'_{sr} \bmod p$\\
            \label{CalSRParmas}
            \textit{sr-ConfigIndex}  $\xleftarrow{}$ \textbf{lookup-tbl}(\textit{subfrm-off}, 3GPP\_36.213~T.10.1.5-1)\\
            \textit{sr-PUCCH-ResourceIndex} $\xleftarrow{}$ $sr$.\textit{sr-PUCCH-ResourceIndex}\\
            \textbf{process} $sr$\\
        }
    \Return{(\textit{sr-ConfigIndex},\textit{sr-PUCCH-ResourceIndex})}
}
\end{algorithm}

\section{Related Work}\label{related-work}
\noindent\textbf{Mobile-relay attacks.} Rupprecht et al.~\cite{rupprecht-breaking-2019} proposes the concept of mobile-relay and demonstrates an attack that redirects the victim's DNS traffic to an attacker controlled server. Then Yang et al.~\cite{yang2019hiding} points out limitations of the relay adversary which must know the radio resource session parameters, which are set up by the eNodeB using encrypted RRC messages. While we use a similar type of adversary as the one proposed by Rupprecht et al.~\cite{rupprecht-breaking-2019}, we are not affected by the shortcomings pointed out by Yang et al.~\cite{yang2019hiding} as we introduce an efficient physical layer parameter guessing procedure which increases the stability of radio connections and makes the mobile-relay undetectable. Furthermore, while the attacks proposed in Rupprecht et al.~\cite{rupprecht-breaking-2019} focus on IP traffic tampering which is being mitigated with the inclusion of integrity protection mechanisms in 5G standards, we show several privacy-related vulnerabilities which remain unmitigated by the above-mentioned standard extensions.

\vspace{5pt}
\noindent\textbf{VoLTE traffic analysis attacks.} A VoLTE attack is proposed by Rupprecht et al.~\cite{rupprecht-call-2020} which exploits the key-stream reuse implementation vulnerability to decrypt voice data transmitted in LTE networks. In this paper we present a different category of \textit{privacy} attack that does not depend on implementation flaws. Our attacks remain applicable even if secure protocols such as Secure Real-time Transport Protocol (SRTP) are deployed or the vulnerability is fixed. We further argue this work has a limitation that requires the malicious call and the victim call must have similar conversion activities. Otherwise, because of the Voice Activity Detection (VAD), the \textit{count} and \textit{length} in PDCP would be different which results in the key-stream becoming different. Our attack does not rely on this assumption.

Kim et al.~\cite{kim2015breaking} analyze the early VoLTE service and find several vulnerabilities caused by weak security policies. Our work focuses on recovering victims' VoLTE logs from the encrypted VoLTE traffic transferred over-the-air. Based on our observation, the vulnerabilities mentioned by Kim et al.~\cite{kim2015breaking}  have been patched nowadays. Lu et al.~\cite{lu2020ghost} also analyze VoLTE services and find several vulnerabilities that could be used to launch session hijacking, DoS and call information leakage. However, they require a stronger attacker model which requires the adversary to be able to obtain the IPsec tunnel keys by installing a malicious application on the victim’s rooted phone. The information leakage observed by them is similar to ours, however, our method of obtaining it requires a weaker adversary and is thus more dangerous.

Xie et al.~\cite{8433136} analyze the Voice Over WiFi (VoWiFi) protocol by looking at the characteristics of plaintext IPsec traffic collected on a malicious AP used to monitor the victim's activity. Our analysis extends on this by analysing the significantly more complex case of VoLTE and VoNR which requires traffic capturing from the LTE/5G encrypted radio link. Here the traffic is encrypted and/or integrity protected using a combination of layers that are part of both IPsec and LTE/5G (i.e. EEA2/EIA2). 

Finally, Kohls et al.~\cite{kohls-lost-2019} and Bae et al.~\cite{277230} analyse the user-plane Internet destined traffic for the purposes of launching fingerprint attacks. Traffic analysis techniques applicable to encrypted Internet traffic are, however, not directly applicable to VoLTE traffic analysis given that voice exchanges are contained exclusively within the carrier network and the traffic is significantly more uniform. To the best of our knowledge, we present the first study which enables the recovery of VoLTE activities by analysing encrypted PDCP packets.

\vspace{5pt}
\noindent\textbf{Identity linking attacks.} Collecting the victim's identifiers (e.g., M-TMSI, SUCI, IMSI) and linking them to the victim's real-life identifiers (e.g., phone number) is the first step to launch more powerful attacks such as location tracking. To collect victim's identifiers, False Base-Station (FBS) attacks have been proposed. These FBS rely on overpowering legitimate signals to attract victim UEs to connect to them instead of legitimate towers. With the addition of mutual-authentication capabilities in 3G/4G and 5G these types of attacks became easily detectable despite some still existing protocol limitations such as the ones outlined by Chlosta et al.~\cite{chlosta20215g} which found that it is still possible to trace the location of the victims using the SUCI in 5G networks.

More recently, Erni et al.~\cite{arxiv.2106.05039} proposes stronger attacks such as \textit{signal overshadowing} which injects \textit{Attach/Service Request} messages in the uplink direction to collect IMSIs. This attack, while able to circumvent the mutual-authentication protections, is still detectable as it causes an observable \textit{Security Command Reject} failure at UE. In this paper, we introduce a method which allows \textit{Attach/Service Request} message tampering without causing a \textit{Security Command Reject} failure. 

The attacks we proposed are also more efficient than \textit{Paging} based attacks, which are commonly used to link victim's identities to real-life identities. As these attacks rely on broadcast messages they normally require (1) several messages to correctly identify a victim from multiple response sets ~\cite{kune2012location,shaik-practical-2015}, and (2) that the victim UE is in the RRC\_IDLE state when the adversary sends the paging message. This further complicates the attack, as the switch from the Paging state to the RRC\_IDLE state takes at least 20s. In contrast, our identity mapping method only requires a single VoLTE \textit{Invite} message to the victim.
\newpage

\begin{table*}[!hp]
    \centering
    \resizebox{\textwidth}{!}{
    \begin{threeparttable}
        \begin{tabular}{l|c|c|c|c|c|c|c|c}
            \hline
            \multirow{3}{*}{VoLTE Signalling} & \multicolumn{6}{c|}{Carrier1} & \multicolumn{2}{c}{Carrier2}\\
            \cline{2-9}
            & \multicolumn{2}{c|}{S7} & \multicolumn{2}{c|}{S8} & \multicolumn{2}{c|}{iPhone11} & \multicolumn{2}{c}{iPhone11}\\
            \cline{2-9}
            & Uplink & Downlink & Uplink & Downlink & Uplink & Downlink & Uplink & Downlink \\
            \hline
            \textit{Invite} & 
                $2479\pm0$\tnote{1} & $2358\pm8$ 
                & $2494\pm0$ & $2357\pm5$
                & $2323\pm0$ & $2371\pm6$ 
                & $2275\pm0$ & $[2219\pm2,2000\pm0]$ \\
            \textit{100 Trying (Invite)} & 
                $338\pm1$ & $445\pm0$ 
                & $336\pm0$ & $445\pm0$
                & - & $409\pm0$ 
                & - & $378\pm0$ \\
            \textit{183 Session Process} &
                $1437\pm1$ & $[1624\pm2,1417\pm1]$ 
                & $1435\pm4$ & $[1623\pm4,1417\pm3]$
                & - & $[1585\pm3,1379\pm3]$ 
                & $1672\pm3$ & $[1519\pm3,852\pm2]$ \\
            \textit{Pack} &
                $1126\pm4$ & $818\pm1$ 
                & $1128\pm2$ & $817\pm2$ 
                & $1229\pm2$ & - 
                & $1174\pm2$ & $[517\pm2,1112\pm0]$ \\
            \textit{200 OK (Pack)} & 
                $715\pm1$ & $[1001\pm4,838\pm0]$ 
                & $713\pm1$ & $[1002\pm2,838\pm0]$ 
                & - & $[962\pm2,802\pm2]$ 
                & $[557\pm2,1234\pm2]$ & $533\pm2$ \\
            \textit{180 Ring (Invite)} &
                $926\pm2$ & $[868\pm2,843\pm1]$ 
                & $894\pm2$ & $[996\pm2,1172\pm2,1159\pm2]$ 
                & $877\pm3$ & $[1199\pm10,1032\pm2]$ 
                & $876\pm4$ & $[1205\pm11,1032\pm0]$ \\
            \textit{486 Busy Here} & 
                $878\pm3$ & - 
                & $878\pm2$ & - 
                & - & - 
                & - & - \\
            \textit{Cancel} & 
                $[639\pm1,986\pm0]$\tnote{2} & $[462\pm0,652\pm1]$ 
                & $[637\pm1,988\pm0]$ & $[462\pm0,650\pm1]$ 
                & $1015\pm0$ & $426\pm0$ 
                & $907\pm0$ & - \\
            \textit{200 OK (Invite)} &
                $[996\pm+1,1435\pm2]$ & $[1086\pm2,1249\pm2]$
                & $994\pm4$ & $[1085\pm0,1249\pm2,1640\pm4]$ 
                & $1365\pm1$ & $[1049\pm2,1212\pm2,1603\pm2]$ 
                & $980\pm2$ & $1140\pm2$ \\
            \textit{ACK (200 OK (Invite))} & 
                $1026\pm4$ & $745\pm1$
                & $1029\pm2$ & $745\pm1$ 
                & $1208\pm2$ & $745\pm1$ 
                & $1152\pm2$ & $527\pm2$ \\
            \textit{487 Request Terminated} & 
                $888\pm2$ & $478\pm0$ 
                & $886\pm2$ & $478\pm0$ 
                & - & $442\pm0$ 
                & - & $563\pm1$ \\
            \textit{ACK (487 ...)} & 
                $672\pm2$ & $392\pm1$ 
                & $672\pm0$ & $390\pm1$
                & $978\pm0$ & - 
                & $916\pm1$ & - \\
            \textit{Bye} & 
                $1104\pm2$ & -
                & $1106\pm2$ & - 
                & $1307\pm6$ & $564\pm1$ 
                & $[1200\pm1,1258\pm1]$ & $1025\pm2$ \\
            \textit{200 OK (Bye)} &
                -\tnote{3} & $459\pm0$ 
                & - & $459\pm0$ 
                & $744\pm1$ & $[402\pm0,423\pm0]$ 
                & $770\pm2$ & $[940\pm1,991\pm2]$ \\
            \textit{Update} & 
                - & $1043\pm1$ 
                & - & -
                & - & - 
                $1805\pm2$ & $1278\pm2$ \\
            \textit{200 OK (Update)} & 
                $1334\pm1$ & - 
                & - & - 
                & - & - 
                & $1460\pm2$ & $1258\pm2$ \\
            \textit{Options} & 
                - & -
                & - & - 
                & - & - 
                & - & $644\pm1$ \\
            \textit{200 OK (Options)} & 
                - & -
                & - & - 
                & - & - 
                & $586\pm1$ & - \\
            \textit{486 Call Rejected By User (Invite)} &
                - & -
                & - & - 
                & $909\pm2$ & - 
                & $939\pm2$ & - \\
            \hline
        \end{tabular}
        \begin{tablenotes}
            \item[1] the observed length is located between $2497-0$ and $2497+0$ bytes.
            \item[2] the observed length is either between $638$ to $640$ bytes or exactly $986$ byes.
            \item[3] this message was not observed.
        \end{tablenotes}
    \end{threeparttable}
    }
    \vspace{5pt}
    \caption{The VoLTE message size (in bytes) for Samsung S7, S8 and iPhone with Carrier1 and iPhone with Carrier2. The size of each signalling type is stable with a small variance. For the downlink messages, the size of each type is quite similar though the UE is different within the same provider. For uplink messages, the size is relevant to phone brands and providers.}
    \label{tab:signal_len_map}
\end{table*}

\end{document}